\documentclass[12pt]{article}
\usepackage{epsfig}
\usepackage{amssymb}

\def\a{\alpha}
\def\b{\beta}

\def\d{\delta}
\def\e{\epsilon}

\def\g{\gamma}

\def\k{\kappa}

\def\n{\nu}
\def\o{\omega}
\def\p{\pi}

\def\s{\sigma}

\def\z{\zeta}
\def\D{\Delta}

\def\G{\Gamma}

\def\O{\Omega}

\def\ve{\varepsilon}

\def\ri{\rm i}
\def\rf{\rm f}

\def\dg{\dagger}                                     
\def\mt{\widetilde{m}_1}                  
\def\mb{\overline{m}}
\def\th{\tilde{h}}
\def\VEV#1{\left\langle #1\right\rangle}        
\def\beq{\begin{equation}}
\def\eeq{\end{equation}}

\def\bea{\begin{eqnarray}}
\def\eea{\end{eqnarray}}
\def\NO{\nonumber}
\def\Bar#1{\overline{#1}}

\def\pl#1#2#3{Phys.~Lett.~{\bf B {#1}} ({#2}) #3}
\def\np#1#2#3{Nucl.~Phys.~{\bf B {#1}} ({#2}) #3}
\def\prl#1#2#3{Phys.~Rev.~Lett.~{\bf #1} ({#2}) #3}
\def\pr#1#2#3{Phys.~Rev.~{\bf D {#1}} ({#2}) #3}

\begin{document}
\date{}
\title{ 
{\normalsize
\mbox{ }\hfill
\begin{minipage}{4cm}   
DESY 03-001\\
UAB-FT-540 \\
CERN-TH/2003-016
\end{minipage}}\\
\vspace{1cm}
\bf The Neutrino Mass Window\\ for Baryogenesis}
\author{W.~Buchm\"uller \\
{\it Deutsches Elektronen-Synchrotron DESY, 22603 Hamburg, Germany}\\[2ex]
P.~Di Bari \\
{\it IFAE, Universitat Aut{\`o}noma de Barcelona,}\\
{\it 08193 Bellaterra (Barcelona), Spain}\\[2ex]
M.~Pl\"umacher\\
{\it Theory Division, CERN, 1211 Geneva 23, Switzerland}
}
\maketitle

\thispagestyle{empty}


\begin{abstract}

\noindent
Interactions of heavy Majorana neutrinos in the thermal phase of the early universe 
may be the origin of the cosmological matter-antimatter asymmetry. This mechanism
of baryogenesis implies stringent constraints on light and heavy Majorana neutrino
masses. We derive an improved upper bound on the $C\!P$ asymmetry in heavy neutrino
decays which, together with the kinetic equations, yields an upper bound on all
light neutrino masses of 0.1~eV. Lepton number changing processes at temperatures
above the temperature $T_B$ of baryogenesis can erase other, pre-existing 
contributions to the baryon asymmetry. We find that these washout processes become
very efficient if the effective neutrino mass $\mt$ is larger than 
$m_* \simeq 10^{-3}$~eV. All memory of the initial conditions is then erased.
Hence, for neutrino masses in the range from 
$\sqrt{\D m^2_{\rm sol}} \simeq 8\times 10^{-3}$~eV to 
$\sqrt{\D m^2_{\rm atm}} \simeq 5\times 10^{-2}$~eV, which is suggested by neutrino
oscillations, leptogenesis emerges as the unique source of the cosmological
matter-antimatter asymmetry. 
\end{abstract}
\newpage

\section{Introduction}

The explanation of the cosmological baryon asymmetry is a challenge for particle
physics and cosmology. In an expanding universe, which leads to departures from 
thermal equilibrium, $C$, $C\!P$ and baryon number violating interactions of quarks 
and leptons can generate dynamically a baryon asymmetry \cite{sak67}.  
The possible realization of these conditions has first been studied in detail in
the context of grand unified theories \cite{yos78,ttx79}.

The picture of baryogenesis is significantly changed by the fact that already in
the standard model of particle physics baryon ($B$) and lepton ($L$) number are not 
conserved due to quantum effects \cite{tho76}. The corresponding non-perturbative
$\D B=3$ and $\D L=3$ processes are strongly suppressed at zero temperature.
However, at temperatures above the critical temperature $T_{EW}$ of the electroweak 
transition they are in thermal equilibrium \cite{krs85} and only the difference
$B-L$ is effectively conserved.

During the past years data on atmospheric and solar neutrinos have provided strong
evidence for neutrino masses and mixings. In the seesaw mechanism \cite{yan79} the
smallness of these neutrino masses $m_\n$ is explained by the mixing $m_D$ of the
left-handed neutrinos with heavy Majorana neutrinos of mass $M$, which yields the
light neutrino mass matrix
\begin{equation}
m_\n = - m_D {1\over M} m_D^T\;.
\end{equation}
Since $m_D ={\cal O}(v)$, where $v \simeq 174$~GeV is the electroweak scale, and 
$M \gg v$, the neutrino masses $m_\n$ are suppressed compared to quark and charged 
lepton masses. $C\!P$ violating interactions of the heavy Majorana neutrinos can 
give rise to a lepton asymmetry and, via the $\D B=3$ and $\D L=3$ sphaleron 
processes, to a related baryon asymmetry. This is the simple and elegant leptogenesis
mechanism \cite{fy86}.

Leptogenesis is a non-equilibrium process which takes place at temperatures 
$T \sim M_1$. For a decay width small compared to the Hubble parameter,
$\G_1(T) < H(T)$, heavy neutrinos are out of thermal equilibrium, otherwise they
are in thermal equilibrium. A rough estimate of the borderline between the two
regimes is given by $\G_1 = H(M_1)$ (cf.~\cite{kt90}). This is equivalent
to the condition that the effective neutrino mass $\mt = (m_D^\dg m_D)_{11}/M_1$ 
equals the `equilibrium neutrino mass'
\begin{equation}\label{mequ}
m_* = {16\p^{5/2}\over 3\sqrt{5}} g_*^{1/2} {v^2\over M_{pl}} 
\simeq 10^{-3}~\mbox{eV}\;,
\end{equation}
where we have used $M_{pl} = 1.2\times 10^{19}$~GeV and $g_* = 434/4$ as effective
number of degrees of freedom. For $\mt > m_*$ ( $\mt < m_*$) the heavy neutrinos of 
type $N_1$ are in (out of) thermal equilibrium at $T=M_1$. 

It is very remarkable that the equilibrium neutrino mass $m_*$ is close to the
neutrino masses suggested by neutrino oscillations,
$\sqrt{\D m^2_{\rm sol}} \simeq 8\times 10^{-3}$~eV and 
$\sqrt{\D m^2_{\rm atm}} \simeq 5\times 10^{-2}$~eV.
This suggests that it may be possible to understand the cosmological baryon 
asymmetry via leptogenesis as a process close to thermal equilibrium. Ideally,
$\D L=1$ and $\D L=2$ processes would be strong enough at temperatures above $M_1$
to keep the heavy neutrinos in thermal equilibrium and weak enough to allow
the generation of an asymmetry at temperatures below $M_1$.  

An analysis of solutions of the Boltzmann equations shows that this is indeed the 
case if light and heavy neutrino masses lie in an appropriate mass range.
In general, the final baryon asymmetry is the result of a competition between
production processes and washout processes which tend to erase any generated
asymmetry. Unless the heavy Majorana neutrinos are partially degenerate,
$M_{2,3}-M_1 \leq M_1$, the dominant processes are decays and inverse decays of  
$N_1$ and the usual off-shell $\D L=1$ and $\D L=2$ scatterings. The final baryon 
asymmetry then depends on just four parameters \cite{bdp021} : the mass $M_1$ of 
$N_1$, the $C\!P$ asymmetry $\ve_1$ in $N_1$ decays, the effective neutrino mass 
$\mt$ and, finally, the sum of all neutrino masses squared, 
$\mb^2 = m_1^2 + m_2^2 + m_3^2$, which controls an important
class of washout processes. Together with the two mass squared differences 
$\D m^2_{\rm atm}$ and $\D m^2_{\rm sol}$, the sum $\mb^2$ determines all neutrino 
masses. Using an upper bound on the $C\!P$ asymmetry $\ve_1$ \cite{hmy02,di02}, 
an upper bound on all light neutrino masses of 0.2~eV has recently been derived 
\cite{bdp022}.

In this paper we extend the previous analysis in two directions. We derive an 
improved upper bound on the $C\!P$ asymmetry which leads to a more stringent
upper bound on light neutrino masses. In addition, we study in detail the washout
of a pre-existing $B-L$ asymmetry, which yields a lower bound on the effective
neutrino mass $\mt$. In this way we obtain a window of neutrino masses for which
leptogenesis can explain the observed cosmological baryon asymmetry, independent
of initial conditions.     

The paper is organized as follows. In Section~2 we derive an improved upper bound
on the $C\!P$ asymmetry $\ve_1$ and illustrate how it can be saturated for
specific neutrino mass matrices. Theoretical expectations for the range of 
neutrino masses are discussed in Section~3. In Section~4 we then derive upper
bounds on the light neutrino masses in the cases of normal and inverted hierarchy,
and we discuss the stability of these bounds.
Section~5 deals with the washout of a large initial $B-L$ asymmetry, and a summary
of our results is given in Section~6.

\section{Bounds on the CP asymmetry}

Given the masses of heavy and light Majorana neutrinos the $C\!P$ asymmetry
$\ve_1$ in the decays of $N_1$, the lightest of the heavy neutrinos,
satisfies an upper bound \cite{hmy02,di02}. In the following we shall study under 
which conditions this upper bound is saturated and how it depends on the
effective neutrino mass $\mt$ which plays an important role in the 
thermodynamic process of leptogenesis.

The standard model with right-handed neutrinos is described by the lagrangian,
\begin{equation}
{\cal L}_m 
= h_{ij}\Bar{l}_{Li}\n_{Rj}\phi   
 + {1\over 2} M_{ij}\Bar{\n}^c_{Ri}\n_{Rj} + h.c.\;,
\end{equation}
where $M$ is the Majorana mass matrix of the right-handed neutrinos, and the
Yukawa couplings $h$ yield the Dirac neutrino mass matrix $m_D = h v$ 
after spontaneous symmetry breaking, $v = \VEV {\phi}$. We work in the
mass eigenstate basis of the right-handed neutrinos where $M$ is diagonal
with real and positive eigenvalues $M_1 \leq M_2 \leq M_3$. The seesaw 
mechanism \cite{yan79} then yields the light neutrino mass matrix
\begin{equation}\label{seesaw}
m_{\n}= - m_D{1\over M}m_D^T  \;, 
\end{equation}
which can be diagonalized by a unitary matrix $U^{(\n)}$,
\begin{equation}\label{mdiag}
U^{(\n)\dg} m_{\n} U^{(\n)*}  = - \left(\begin{array}{ccc}
    m_1  & 0  & 0\\
    0   &  m_2   & 0  \\
    0  & 0    & m_3
    \end{array}\right) \;,
\end{equation}
with real and positive eigenvalues satisfying $m_1 \leq m_2 \leq m_3$.

It is convenient to work in a basis where also the light neutrino mass matrix
is diagonal. In this basis the Yukawa couplings are
\begin{equation}\label{yeigen}
\tilde{h} = U^{(\n)\dg} h\;.
\end{equation}
As a consequence of the seesaw formula the matrix $\O$,
\begin{equation}\label{ortho}
\O_{ij} = {v\over \sqrt{m_iM_j}}\th_{ij}\;,
\end{equation}
is orthogonal, $\O \O^T = \O^T \O = I$ \cite{ci01}. It is then easy to show
that the $C\!P$ asymmetry $\ve_1$ \cite{fps95}-\cite{bp98} is given by (cf., e.g., 
\cite{bdp021})
\begin{equation}\label{nice}
\ve_1 = {3\over 16\p} {M_1 \over v^2} \sum_{i \neq 1} {\D m_{i1}^2\over m_i}
{\mbox{Im}\left(\th^2_{i1}\right) \over 
\left(\th^\dg \th\right)_{11}}\; ,
\end{equation} 
where $\D m_{i1}^2 = m_i^2 - m_1^2$.

The $C\!P$ asymmetry $\ve_1$ is bounded by the maximal asymmetry $\ve_1^{max}$
\cite{bdp022}, 
\begin{equation}\label{emax}
|\ve_1| \leq \ve_1^{max} = {3 \over 16 \p} {M_1\over v^2} 
{(\D m^2_{atm} + \D m^2_{sol})\over m_3} \;.
\end{equation}
As we will now show, this bound holds for arbitrary values of $m_2$, i.e. for
normal and for inverted hierarchy, and it is saturated in the limit 
$m_1 \rightarrow 0$. 

Consider the normalized Yukawa couplings
\begin{equation}
z_i = {\th^2_{i1} \over (\th^\dg\th)_{11}} = x_i + i y_i\;,
\end{equation}
with
\begin{equation}\label{constraint}
0 \leq |z_i| \leq 1\; , \quad \sum_i |z_i| = 1\;.
\end{equation} 
The orthogonality condition $(\O^T \O)_{11} = 1$ yields the additional 
constraint
\begin{equation}
\sum_i {\mt \over m_i} z_i = 1\;.
\end{equation}
In the new variables the $C\!P$ asymmetry reads 
\begin{equation}\label{nicer}
\ve_1 = {3\over 16\p} {M_1 \over v^2} \left({\D m_{21}^2\over m_2} y_2 + 
{\D m_{31}^2\over m_3} y_3\right)\; .
\end{equation} 
Since $m_3 > m_2$, one also has $\D m^2_{31}/m_3 > \D m^2_{21}/m_2$. This suggests
that the maximal $C\!P$ asymmetry is reached for maximal $y_3$.

Suppose now that $1-y_3 = {\cal O}(\e)$. Because of 
Eqs.~(\ref{constraint}) this implies $y_2$, $y_1$ and all $x_i$ have to vanish
in the limit $\e \rightarrow 0$. The orthogonality condition 
$(\O^T \O)_{11} = 1$ yields
\begin{equation}\label{ima}
{y_1 \over m_1} + {y_2 \over m_2} + {y_3 \over m_3} = 0\;,
\end{equation}
\begin{equation}\label{rea}
{\mt \over m_1}x_1 + {\mt \over m_2}x_2 + {\mt \over m_3}x_3 = 1\;.
\end{equation}
Since $m_2 > 0$, these conditions are satisfied for maximal $y_3$, if
$y_2 = x_2 = x_3 = 0$ and
\begin{equation}
m_1 \;,\; y_1 \propto \e\;,
\end{equation}
\begin{equation}
\mt \propto \e^a\;, \quad x_1 \propto \e^{1-a}\;, \quad 0 \leq a < 1\;.
\end{equation}
Note that in the limit $\e \rightarrow 0$, $N_1$ couples only to $l_3\phi$.
For $a > 0$, $N_1$ decouples completely, since $\th^2_{i1} = 
(\th^\dg\th)_{11} z_i$ and $(\th^\dg\th)_{11} \propto \mt$. 

An explicit example, which illustrates this saturation of the $C\!P$ bound, is
given by the following orthogonal matrix,
\begin{equation} 
\O = \left(\begin{array}{ccc}
    A  & 0  & -B \\
    0  & 1  & 0  \\
    B  & 0  & A
    \end{array}\right) \;,
\end{equation}
with
\begin{equation}
B^2 = i{v^2\over m_3 M_1} b \e^a\;, \quad A^2 = 1 - B^2\;, \quad b>0\;.
\end{equation}
The corresponding Yukawa couplings squared are
\begin{equation}
\left(\th^2_{i1}\right) = 
\left({m_1 M_1\over v^2}-i{m_1\over m_3}b\e^a, 0, ib\e^a\right)\;.
\end{equation}
One obviously has $x_2 = x_3 = y_2 = 0$, and $y_3 \rightarrow 1$,
$x_1, y_1 \rightarrow 0$ in the limit $\e \rightarrow 0$. The matrix
of Yukawa couplings,
\begin{equation}
\th = \left(\begin{array}{ccc}
    \sqrt{{m_1 M_1\over v^2}-i{m_1\over m_3} b\e^a}  & 0  & 
                               \sqrt{i{m_1 M_3\over m_3 M_1}b\e^a} \\
    0  & {\sqrt{m_2M_2}\over v}  & 0  \\
    -\sqrt{ib\e^a}  & 0  & 
                    \sqrt{{m_3 M_3\over v^2}-i{M_3\over M_1}b\e^a}
    \end{array}\right)\;,
\end{equation}
becomes diagonal in the limit $\e \rightarrow 0$ for $a>0$. Hence, in this basis, the
large neutrino mixings are due to the charged lepton mass matrix.

This example illustrates that $\mt$ can be arbitrary in the limit
$m_1 \rightarrow 0$. It approaches $b^2v^2/M_1$ for $a=0$, while it goes to 0 
for $a>0$. Hence, the maximal $C\!P$ asymmetry
(\ref{emax}) can be reached for arbitrary values of $m_2$ and $\mt$. For a
given $C\!P$ asymmetry, the maximal baryon asymmetry is reached in the limit
$\mt \rightarrow 0$, assuming thermal initial $N_1$ abundance. The corresponding, 
model independent lower bound on the heavy neutrino mass $M_1$ was determined in 
\cite{bdp021} to be $M_1 > 4\times 10^8$~GeV. If the Yukawa couplings $\th$ are 
restricted, a more stringent lower bound on $M_1$ can be derived \cite{er02}.

The above discussion can easily be extended to derive the maximal CP asymmetry
in the case of arbitrary $\mt$. Since $m_3 > m_2 > m_1$, one again has 
$x_3 = x_2 = y_2 = 0$. From Eqs.~(\ref{ima}),(\ref{rea}) one then concludes
\begin{equation}
y_1 = - {m_1\over m_3} y_3\;, \quad  x_1 = {m_1\over \mt}\;.
\end{equation} 
Together with the constraint (cf.~(\ref{constraint})),
$\sqrt{x_1^2 + y_1^2} + |y_3| = 1$, these conditions determine $|y_3|$ as
function of $m_1$, $m_3$ and $\mt$. Inserting the result into Eq.~(\ref{nicer})
yields the improved upper bound
\begin{equation}\label{iemax}
\ve_1^{max} = {3\over 16\p}{M_1m_3\over v^2}
\left[1-{m_1\over m_3}\left(1+{m_3^2-m_1^2\over \mt^2}\right)^{1/2}\right]\;.
\end{equation}
For $m_1=0$ the result coincides with the previous bound (\ref{emax}).  
For $0 < m_1 \leq \mt$ the new bound is more stringent. In particular,
$\ve_1^{max} = 0$ for $\mt=m_1$. Note that according to Eq.~(\ref{iemax}) the
only model independent restriction on the effective neutrino mass is 
$\mt \geq m_1$. The improved upper bound on  the $C\!P$ asymmetry implies also a 
bound on the light neutrino masses which is more stringent than the one obtained in 
\cite{bdp022}. This will be discussed in Section~4. 

\section{Range of neutrino masses}

At present we know two mass squared differences for the light neutrinos, which are
deduced from the measurements of solar and atmospheric neutrino fluxes. In 
addition we have information about elements of the mixing matrix $U$ in the
leptonic charged current. Since $U$ could be entirely due to mixings of the
charged leptons, this does not constrain the light neutrino mass matrix in a
model independent way. The light neutrino masses $m_1 < m_2 <m_3$ can be either
quasi-degenerate or hierarchical, with $m_2 - m_1 \ll m_3 - m_2$ (`normal hierarchy')
or $m_3 - m_2 \ll m_2 - m_1$ (`inverted hierarchy'). The best information on the 
absolute neutrino mass scale comes from neutrinoless double $\b$-decay, which yields 
an upper bound on the light Majorana neutrino masses of about 1~eV \cite{gen01,fsv02}. 

A crucial quantity for thermal leptogenesis is the effective neutrino mass $\mt$
which is always larger than $m_1$ \cite{fhy02}, as one easily sees from the 
orthogonality of $\O$ (cf.~(\ref{ortho})),
\begin{eqnarray}\label{mt1lbound}
\mt &=& {v^2\over M_1}\sum_i |\tilde{h}_{i1}^2|\; =\; \sum_i m_i |\O_{i1}^2| \NO\\
 &\geq& m_1 \sum_i |\O_{i1}^2|\; \geq\; m_1 \sum_i {\rm Re}(\O_{i1}^2)\;
 = \; m_1\; .
\end{eqnarray}
As we saw in the previous Section, the maximal $C\!P$ asymmetry is reached for
$m_1 = 0$, such that
$m_2 \simeq \sqrt{\D m^2_{\rm sol}}$ and $m_3 \simeq \sqrt{\D m^2_{\rm atm}}$. 

There is no model independent upper bound on $\mt$. However, if there are no strong
cancelations due to phase relations between different matrix elements, one has
\begin{equation}
\mt \leq m_3 \sum_i |\O_{i1}^2| \sim m_3 |\sum_i \O_{i1}^2| = m_3\; .
\end{equation} 
Hence, the natural range for the effective neutrino mass is $m_1\leq\mt\lesssim m_3$. 
In fact, we are not aware of any neutrino mass model where this is not the case.

It is instructive to examine the range of $\mt$ also in the special case  
$|\ve_1|=\ve_1^{\rm max}$. As we saw in the previous section this case is realized 
for $y_2=x_2=x_3=0$, corresponding to 
${\rm Re} (\O_{21}^2)={\rm Re} (\O_{31}^2)={\rm Im} (\O_{21}^2)=0$.
The orthogonality condition then implies 
${\rm Im}(\O_{11}^2)=-{\rm Im} (\O_{31}^2)$ and ${\rm Re} (\Omega_{11}^2)=1$. Hence,
for maximal $C\!P$ asymmetry one has
\begin{equation}
\mt = m_1\sqrt{1+{\rm Im}(\O_{31}^2)^2} + m_3\,|{\rm Im}(\O_{31}^2)|\; ,
\end{equation}
showing that the value of $\widetilde{m}_1$ is tuned by just one quantity. 
For ${\rm Im}(\O_{31}^2)=0$, one has $\mt=m_1$, while the case $\mt \gg m_3$
corresponds to a fine tuned situation in which 
$|{\rm Im}(\O_{31}^2)|=|{\rm Im}(\O_{11}^2)|\gg {\rm Re}(\O_{11}^2)=1$. 

If the observed large mixing angles in the leptonic charged current originate from
the neutrino mass matrix, which appears natural since their Majorana nature
distinguishes neutrinos from quarks, the masses $m_1$ and $\mt$ are related to
$m_2$ and $m_3$. The seesaw mechanism together with leptogenesis then also 
constrains the heavy Majorana neutrino masses.

Large mixing angles are naturally explained if neutrino masses are quasi-degenerate
\cite{fx02}. One then has $\mt \approx m_1 \approx m_2 \approx m_3 > 0.1$~eV. 
However, as shown in \cite{bdp021,bdp022} and further strengthened in the 
following Section, quasi-degenerate neutrinos are strongly disfavored by thermal 
leptogenesis. A possible exception is the case where also the heavy Majorana 
neutrinos are partially degenerate. One then gets an enhancement of the
$C\!P$ asymmetry which allows one to increase the neutrino masses and still have 
successful leptogenesis. Models with 
$\D M_{21}/M_1 = (M_2 - M_1)/M_1 < 5\times 10^{-2}$ and
$\D M_{21}/M_1 = 5\times 10^{-7}$ have been considered in refs.~\cite{ery02} and
\cite{bgj02}, respectively. Note, however, that in these examples the light 
neutrino masses are not quasi-degenerate. We shall pursue this case further in
Section~4.3.

The neutrino mass pattern with inverted hierarchy has also received much attention
in the literature. There is, however, the well known difficulty of this scenario
to fit the large angle MSW solution \cite{bd01,af02}. We also do not know any
model with inverted hierarchy which incorporates successfully leptogenesis, and
we shall therefore not pursue this case further.

We are then left with the case of neutrino masses with normal hierarchy. There are
many neutrino mass models of this type with successful leptogenesis. The mass hierarchy
is usually controlled by a parameter $\e \ll 1$. For the effective neutrino mass one
can then have, for instance, $\mt \sim m_2$ (cf.~\cite{bgj02,bp96}). A simple and 
attractive form of the light neutrino mass matrix, which can account for all data, is 
given by \cite{sy98,ilr98},
\begin{equation}\label{mnu}
m_\n \sim \left(\begin{array}{ccc}
    \e^2  & \e  & \e \\
    \e  & 1  & 1  \\
    \e  & 1  & 1 \end{array}\right) {v^2\over M_3} \;,
\end{equation}
where coefficients ${\cal O}(1)$ have been omitted. This form could follow from 
a U(1) family symmetry \cite{fn79} or a relation between the hierarchies of Dirac
and Majorana neutrino masses \cite{bw01}. In the second case one has
$m_1,m_2 \sim \e m_3$ and $\mt \sim m_3$, which is compatible with leptogenesis. 
The structure of the mass matrix (\ref{mnu}) as well as predictions for the
coefficients ${\cal O}(1)$ can be obtained in seesaw models where the exchange of
two heavy Majorana neutrinos dominates \cite{kin00}. In all these examples the
range of the effective neutrino mass is $m_1 \le \mt \lesssim m_3$. 

Thermal leptogenesis also leads to a lower bound on $M_1$, the smallest of the
heavy neutrino masses \cite{by93,di02}. In the minimal scenario, where the heavy
neutrinos are not degenerate, one obtains the lower bound $M_1 > 4\times 10^8$~GeV
\cite{bdp021}. It is reached for maximal $C\!P$ asymmetry ($m_1=0$), minimal
washout ($\mt \rightarrow 0$), and assuming thermal initial $N_1$ abundance. The 
bound becomes more stringent for restricted patterns of mass matrices \cite{er02}. 
It can be relaxed if the heavy neutrinos are partially degenerate \cite{pil99,ery02,bgj02}. 

\section{Improved upper bounds on neutrino masses}

\subsection{Maximal asymmetry and CMB constraint}

It is useful to recast the maximal $C\!P$ asymmetry (\ref{iemax}) in the following way,
\begin{equation}
\ve_1^{\rm max}(M_1,\mt,\mb) = 10^{-6}\,\left(M_1\over 10^{10} {\rm GeV} \right)\,
{m_{\rm atm}\over m_0}\,\,\b(\mt,\mb)\; ,
\end{equation}
where 
$m_{\rm atm}= \sqrt{\Delta m^2_{\rm atm}+\Delta m^2_{\rm sol}}$,
$m_0=(16\,\pi/3)\,(v^2/10^{10}\,{\rm GeV})\simeq 0.051\,{\rm eV}$,
and 
\begin{equation}\label{beta}
\b(\mt,\mb) = \,\,{\left(m_3-m_1\,\sqrt{1+{m^2_{\rm atm}\over 
\widetilde{m}_1^2}}\right)\over m_{\rm atm}}
\leq 1\,\,\,\, .
\end{equation}
The maximal value, $\b=1$, is obtained for $m_1=0$. Note, that $m_{\rm atm}=m_0$
for the best fit values extracted from the KamLAND data \cite{kam02},
$\D m^2_{\rm sol}= 6.9\times 10^{-5}\,{\rm eV^2}$, 
and the SuperKamiokande data \cite{sk02},
$\D m^2_{\rm atm}= 2.5\times 10^{-3}\,{\rm eV}^2$. 

We will calculate particle numbers and asymmetries normalized to the number of photons 
per comoving volume before the onset of leptogenesis at $t_{\star}$ \cite{bdp021}. 
For zero initial $B-L$ asymmetry, i.e. $N_{B-L}^{\rm i}=0$, the final $B-L$ asymmetry 
produced by leptogenesis is given by
\begin{equation}\label{NBmLf}
N_{B-L}^{\rm f}=-{3\over 4}\,\varepsilon_1\,\kappa_{\rm f}\;,
\end{equation}
where $\kappa_{\rm f}$ is the `efficiency factor'. In the minimal version of thermal
leptogenesis one considers initial temperatures $T_i \gtrsim M_1$, where $M_1$ is the 
mass of the
lightest heavy neutrino $N_1$. In this case $\kappa_{\rm f}\leq 1$, and the maximal 
value, $\kappa_{\rm f}=1$, is obtained for thermal initial $N_1$ abundance in the 
limit $\mt\rightarrow 0$. The heavy neutrinos $N_1$ then decay fully out of
equilibrium at temperatures well below $M_1$, producing a $B-L$ asymmetry which
survives until today since all washout processes are frozen at temperatures $T \ll M_1$.
 
In the case of general initial conditions and arbitrary values of $\mt$, the efficiency
factor $\kappa_{\rm f}$ has to be calculated by
solving the Boltzmann equations \cite{lut92,plu97,plu98,bcx00,bdp021}, 
\begin{eqnarray}\label{ke}
{dN_{N_1}\over dz} & = & -(D+S)\,(N_{N_1}-N_{N_1}^{\rm eq}) \;, \label{lg1} \\ 
{dN_{B-L}\over dz} & = & -\ve_1\,D\,(N_{N_1}-N_{N_1}^{\rm eq})-W\,N_{B-L} \;,\label{lg2}
\end{eqnarray}
where $z=M_1/T$. There are four classes of processes which contribute to the
different terms in the equations: decays, inverse decays, $\D L=1$ scatterings 
and processes mediated by heavy neutrinos. The first three all modify the 
$N_1$ abundance. Denoting by $H$ the Hubble parameter, $D = \Gamma_D/(H\,z)$ 
accounts for decays and inverse decays, while $S = \Gamma_S/(H\,z)$ represents
the $\D L=1$ scatterings. The decays are also the source term for the generation of 
the $B-L$ asymmetry, the first term in Eq.~(\ref{lg2}), while all the other processes 
contribute to the total washout term $W = \Gamma_W/(H\,z)$ which competes 
with the decay source term. 

We take into account only decays of $N_1$, neglecting the decays of the heavier
neutrinos $N_2$ and $N_3$. These decays produce a $B-L$ asymmetry at temperatures 
higher than $M_1$. As we shall see in Section~5, the washout processes at 
$T\sim M_1$ very efficiently erase any previously generated asymmetry. Even in the 
case of very small mass differences the decays of $N_2$ and $N_3$ do not change 
significantly the bound on the light neutrino masses, which is our main interest.
This will be discussed in Section~4.3.

The baryon to photon number ratio at recombination, $\eta_B$, is simply related
to $N_{B-L}^{\rm f}$ by $\eta_B=(a/f)\,N_{B-L}^{\rm f}$, where $a=28/79$ 
\cite{ks88} is the fraction of $B-L$ asymmetry which is converted into a baryon 
asymmetry by sphaleron processes, and $f=N_{\g}^{\rm rec}/N_{\g}^{\star}=2387/86$ 
accounts for the dilution of the asymmetry due to standard photon production from the
onset of leptogenesis till recombination. $\eta_B^{\rm max}$, the final baryon 
asymmetry produced by leptogenesis with maximal $C\!P$ asymmetry, i.e.
$\ve_1=\ve_1^{\rm max}$), is given by
\begin{equation}
\eta_{B}^{\rm max}\simeq 0.96\times 10^{-2}\,\varepsilon_1^{\rm max}\,\k_{\rm f}\;. 
\end{equation}
This quantity has to be compared with measurements of the CMB experiments 
BOOMerANG \cite{boo02} and DASI \cite{das02},
\begin{equation}\label{CMB}
\eta_{B}^{CMB}=(6.0^{+1.1}_{-0.8})\times 10^{-10} \;.
\end{equation}
The CMB constraint then requires $\eta_B^{\rm max}\geq \eta_{B}^{CMB}$, and we 
will adopt for $\eta_{B}^{CMB}$ the $3\sigma$ lower limit,
$(\eta_{B}^{CMB})_{\rm low}=3.6\times 10^{-10}$. 

In \cite{bdp021} we showed that $\eta_{B}^{\rm max}$ depends just on the three 
parameters $\mt,M_1$ and $\mb$. Thus, for a given value of $\mb$, the CMB constraint 
determines an allowed region in the $(\mt,M_1)$-plane. It was also shown that there 
is an upper bound for $\mb$ above which no allowed region exists. In \cite{bdp022}, 
based on the bound (\ref{emax}) for the $C\!P$ asymmetry, $\mb<0.30\,{\rm eV}$ was
derived as upper bound on the neutrino mass scale. In the following we shall 
study the allowed regions in the $(\mt,M_1)$-plane for different parameters $\mb$ 
using the improved bound on the $C\!P$ asymmetry (\ref{iemax}) and in this way 
determine a new improved bound on $\mb$. 

\subsection{Numerical results} 

The neutrino masses $m_1$ and $m_3$ depend in a different way on $\mb$ in the cases
of normal and inverted hierarchy, respectively. Hence, also the dependence of the
function $\b$ on $\mb$ is different for these two mass patterns. This leads to different
maximal baryon asymmetries $\eta_{B}^{\rm max}$, and therefore to different upper 
bounds on $\mb$, in the two cases which we now study in turn.

For neutrino masses with normal hierarchy one has
\begin{equation} \label{nD32}
m^{\,2}_3-m^{\,2}_2  = \D m^2_{\rm atm}\; , \quad
m^{\,2}_2-m^{\,2}_1  = \D m^2_{\rm sol}\; , \label{nD21} 
\end{equation} 
and the dependence on $\mb$ is given by
\begin{eqnarray} \label{numanor1}
m_3^{\,2} 
&=& {1\over 3}\left(\mb^2 + 2\Delta m^2_{\rm atm} + \Delta m^2_{\rm sol}\right)\;, \\
\label{numanor2}
m_2^{\,2} 
&=& {1\over 3}\left(\mb^2 - \Delta m^2_{\rm atm} + \Delta m^2_{\rm sol}\right)\;, \\
\label{numanor3}
m_1^{\,2} 
&=& {1\over 3}\left(\mb^2 - \Delta m^2_{\rm atm} - 2\Delta m^2_{\rm sol}\right)\;. 
\end{eqnarray}
These relations are plotted in Fig.~1. Note that there is a minimal value of $\mb$, 
corresponding to $m_1=0$, which is given by 
$\mb_{\rm min}=\sqrt{\D m^2_{\rm atm} + 2\D m^2_{\rm sol}} \simeq 0.051\,{\rm eV}$. 

\begin{figure}
\centerline{\psfig{figure=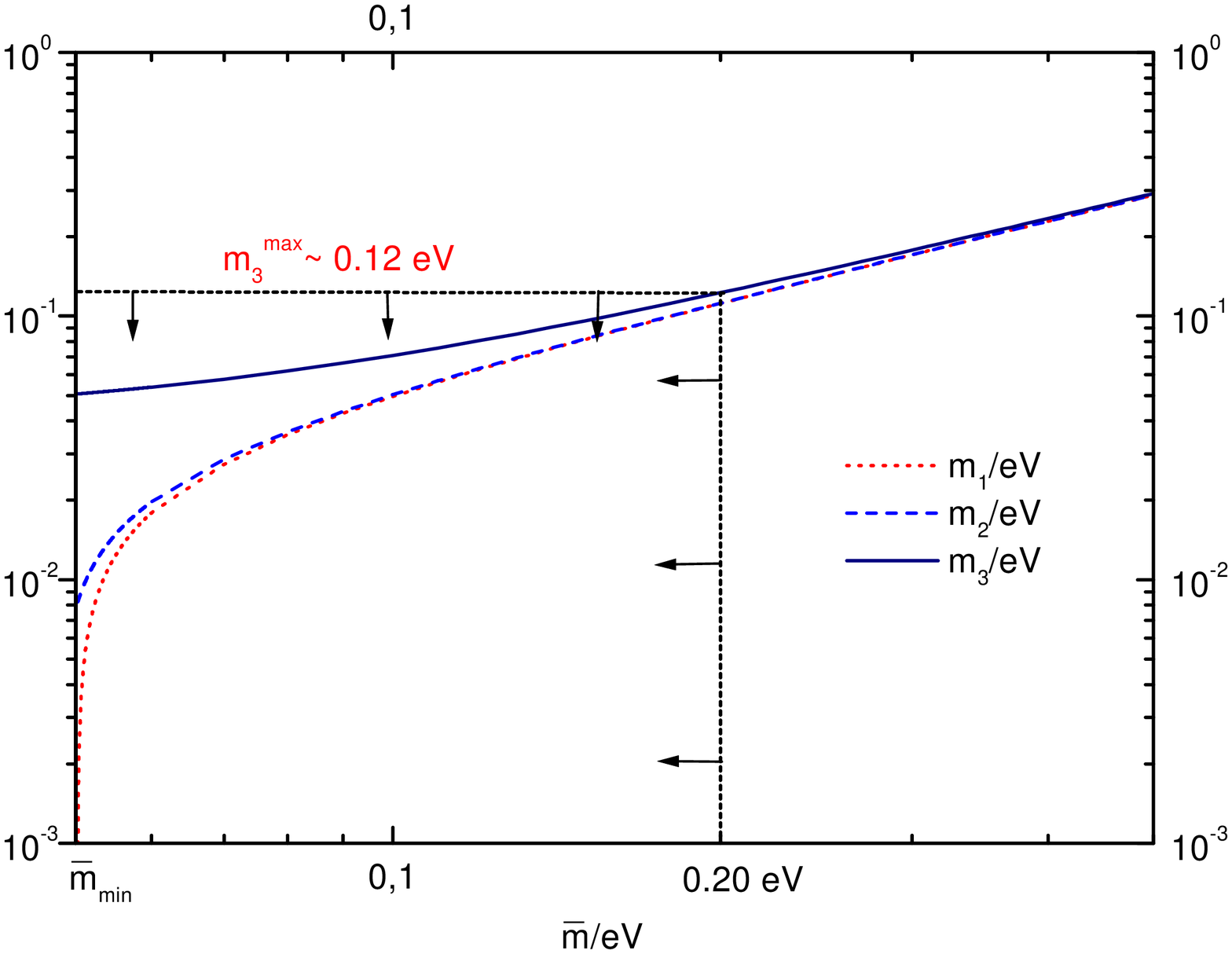,height=88mm,width=13cm}}
\caption{\small Neutrino masses as functions of $\mb$ for 
normal hierarchy (cf.~(\ref{numanor1})-(\ref{numanor3})).}
\centerline{\psfig{file=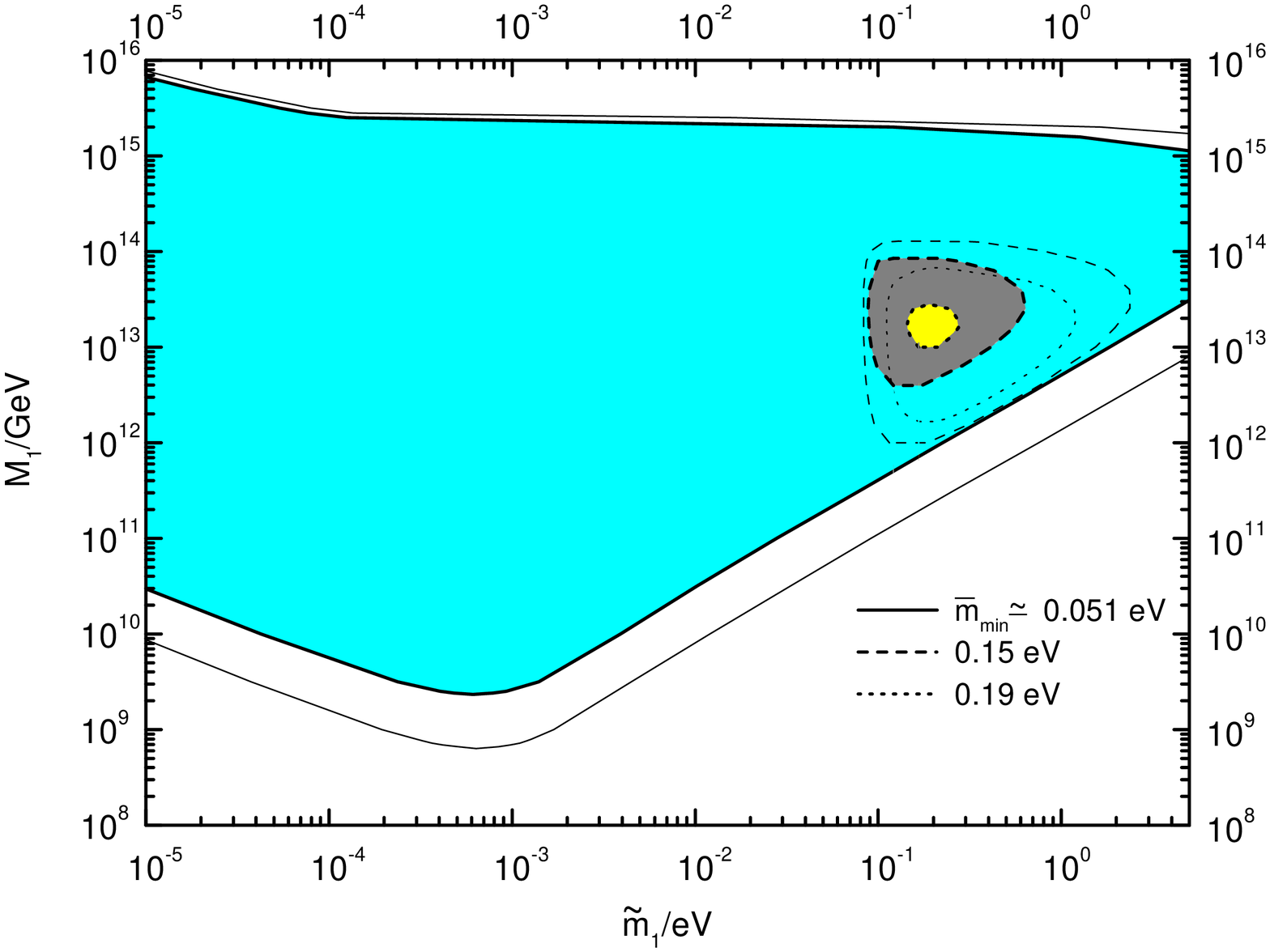,height=87mm,width=13cm}}
\caption{{\small Normal hierarchy case.
Curves, in the ($\mt$-$M_1$)-plane, of constant
$\eta_{B0}^{\rm max}=10^{-10}$ (thin lines) 
and $\eta_{B0}^{\rm max}=3.6\times 10^{-10}$ (thick lines)
for the indicated values of $\mb$.
The filled regions for 
$\eta^{\rm max}_{B0}\geq 3.6\times 10^{-10}$ are the {\it allowed regions}
from CMB. 
There is no allowed region for $\mb=0.20\,{\rm eV}$.}}
\end{figure}

Fig.~2 shows the lines of constant maximal baryon asymmetry 
$\eta_{B}^{\rm max}=(\eta_{B}^{CMB})_{\rm low}$ (thick lines)
and $\eta_{B}^{\rm max}=10^{-10}$ (thin lines) in the $(\widetilde{m}_1,M_1)$-plane
for different choices of $\mb$ and assuming zero initial $N_1$ abundance.
The allowed regions (the filled ones)
correspond to the constraint $\eta_{B}^{\rm max}\geq (\eta_{B}^{CMB})_{\rm low}$. 
The largest allowed region is obtained for $\mb=\mb_{\rm min}$, since
in this case the $C\!P$ asymmetry is maximal, i.e. $\b=1$ for any value of $\mt$, 
and the washout is minimal. Note that a different choice for the initial 
$N_1$ abundance would have affected the final baryon asymmetry only for 
$\mt < m_*$. 
The case of an initial thermal abundance has been studied in \cite{bdp021}.
When $\mb$ increases different effects combine to shrink the allowed region until it 
completely disappears at some value $\mb_{\rm max}$. 

We have determined this value with a numerical uncertainly of $0.01\,{\rm eV}$. 
From Fig.~2 one can see that there is a small allowed region for $\mb=0.19\,{\rm eV}$,
whereas we found no allowed region  for $\mb=0.20\,{\rm eV}$. Hence, the value of 
$\mb_{\rm max}$ is somewhere in between and we can conclude that in the case of normal 
hierarchy,
\begin{equation}\label{boundn}
\mb < 0.20\,{\rm eV}\; .
\end{equation}
Using the relations (\ref{numanor1})-(\ref{numanor3}), one can easily translate this 
bound into upper limits on the individual neutrino masses (cf.~Fig.~1),
\begin{equation}
m_1,m_2 < 0.11\,{\rm eV}\; ,\;\;\; m_3 < 0.12\,{\rm eV}\; .
\end{equation}

The case of an inverted hierarchy of neutrino masses corresponds to 
\begin{equation} \label{iD32}
m^2_3-m^2_2 = \Delta m^2_{\rm sol}\; , \quad
m^2_2-m^2_1 = \Delta m^2_{\rm atm} \label{iD21}\;, 
\end{equation} 
and the relations between the neutrino masses and $\mb$ are 
\begin{eqnarray}\label{numainv1}
m_3^2 &=& {1\over 3}\left(\mb^2 + \D m^2_{\rm atm} + 2\,\D m^2_{\rm sol}\right)\;, \\
m_2^2 &=& {1\over 3}\left(\mb^2 + \D m^2_{\rm atm} - \D m^2_{\rm sol}\right)\;, \\
\label{numainv3}
m_1^2 &=& {1\over 3}\left(\mb^2 - 2\,\D m^2_{\rm atm} - \D m^2_{\rm sol}\right)\;. 
\end{eqnarray}
We have plotted these relations in Fig.~3.
\begin{figure}
\hspace{5mm}
\centerline{\psfig{figure=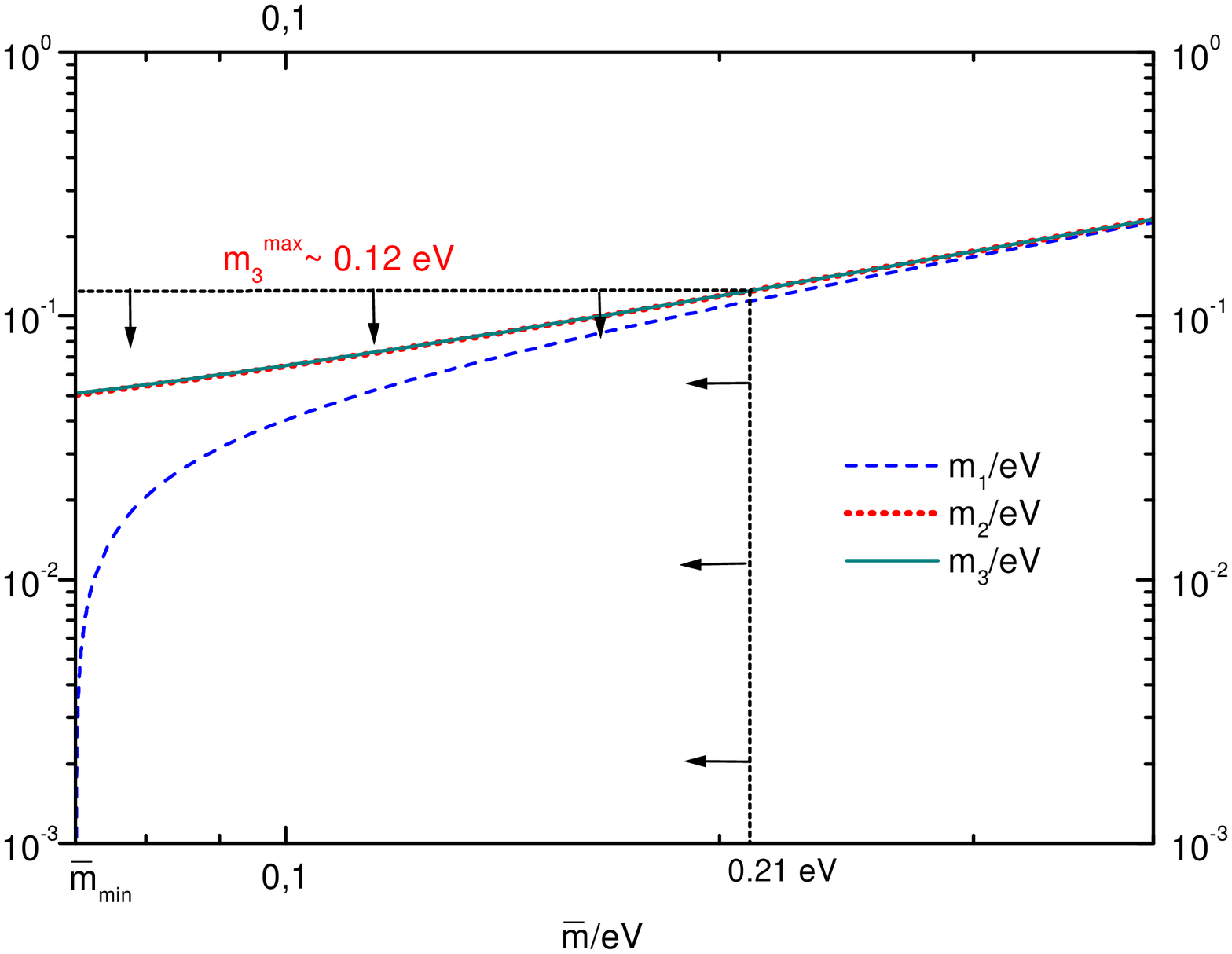,height=88mm,width=13cm}}
\caption{\small Neutrino masses as functions of $\mb$ for
inverted hierarchy (cf.~(\ref{numainv1})-(\ref{numainv3})).}
\centerline{\psfig{file=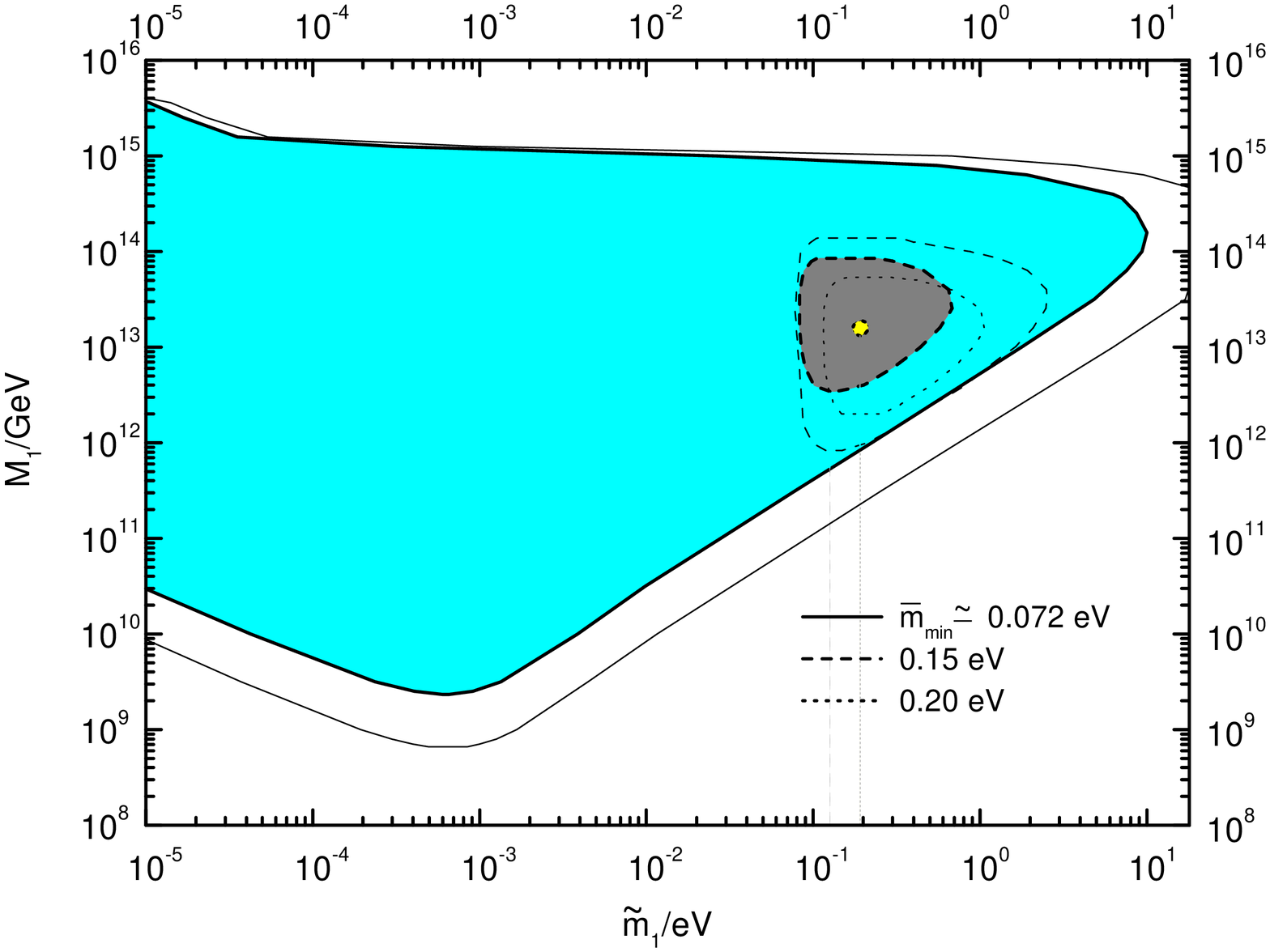,height=88mm,width=13cm}}
\caption{{\small Inverted hierarchy case.
Curves, in the ($\mt$-$M_1$)-plane, of constant
$\eta_{B0}^{\rm max}=10^{-10}$ (thin lines) 
and $\eta_{B0}^{\rm max}=3.6\times 10^{-10}$ (thick lines)
for the indicated values of $\mb$.
The filled regions for 
$\eta^{\rm max}_{B0}\geq 3.6\times 10^{-10}$ are the {\it allowed regions}
from CMB. There is no allowed region for $\mb=0.20\,{\rm eV}$.}}
\end{figure}
The minimal value of $\mb$, corresponding to $m_1=0$, is now 
$\mb_{\rm min}=\sqrt{2\,\D m^2_{\rm atm} + \D m^2_{\rm sol}}\simeq 0.072\,{\rm eV}$.

The curves of constant $\eta_{B}^{\rm max}$ are shown in Fig.~4 for different values of 
$\mb$. The largest allowed region  is again obtained for $\mb=\mb_{\rm min}$.
One can see that this time there is a tiny allowed region for $\mb=0.20\,{\rm eV}$
and no allowed region for $\mb=0.21\,{\rm eV}$. Therefore, in the case
of inverted hierarchy the upper bound is slightly relaxed, 
\begin{equation}\label{boundi}
\mb< 0.21\,{\rm eV}\; .
\end{equation}
Using the relations (\ref{numainv1})-(\ref{numainv3}) one can 
again translate the bound on $\mb$ into bounds on the individual neutrino masses,
\begin{equation}
m_1< 0.11\,{\rm eV}\, ,\;\;\;m_2,m_3<0.12\,{\rm eV}\; .
\end{equation}

Let us now discuss the different effects which combine to shrink the allowed region 
when the absolute neutrino mass scale $\mb$ increases, thus yielding the upper bound. 
The first effect is that away from the hierarchical neutrino case, for 
$\mb>\mb_{\rm min}$ and $m_1>0$, the maximal $C\!P$ asymmetry  reduces considerably. 
This can be seen in terms of the function  $\b$ (cf.~(\ref{beta})) which is
conveniently expressed in the form
\begin{equation}
\b = \b_{\rm max}(\mb)\,f(\mt,\mb)\; .
\end{equation}
The first factor, $\b_{\rm max}=(m_3-m_1)/m_{\rm atm}=m_{\rm atm}/(m_3+m_1)$, 
is the maximal value of $\b$ for fixed $\mb$; $\b_{\rm max}$ decreases
$\propto 1/\mb$ for $\mb \gg \mb_{\rm min}$ (cf.~Fig.~5).
\begin{figure}
\centerline{\psfig{file=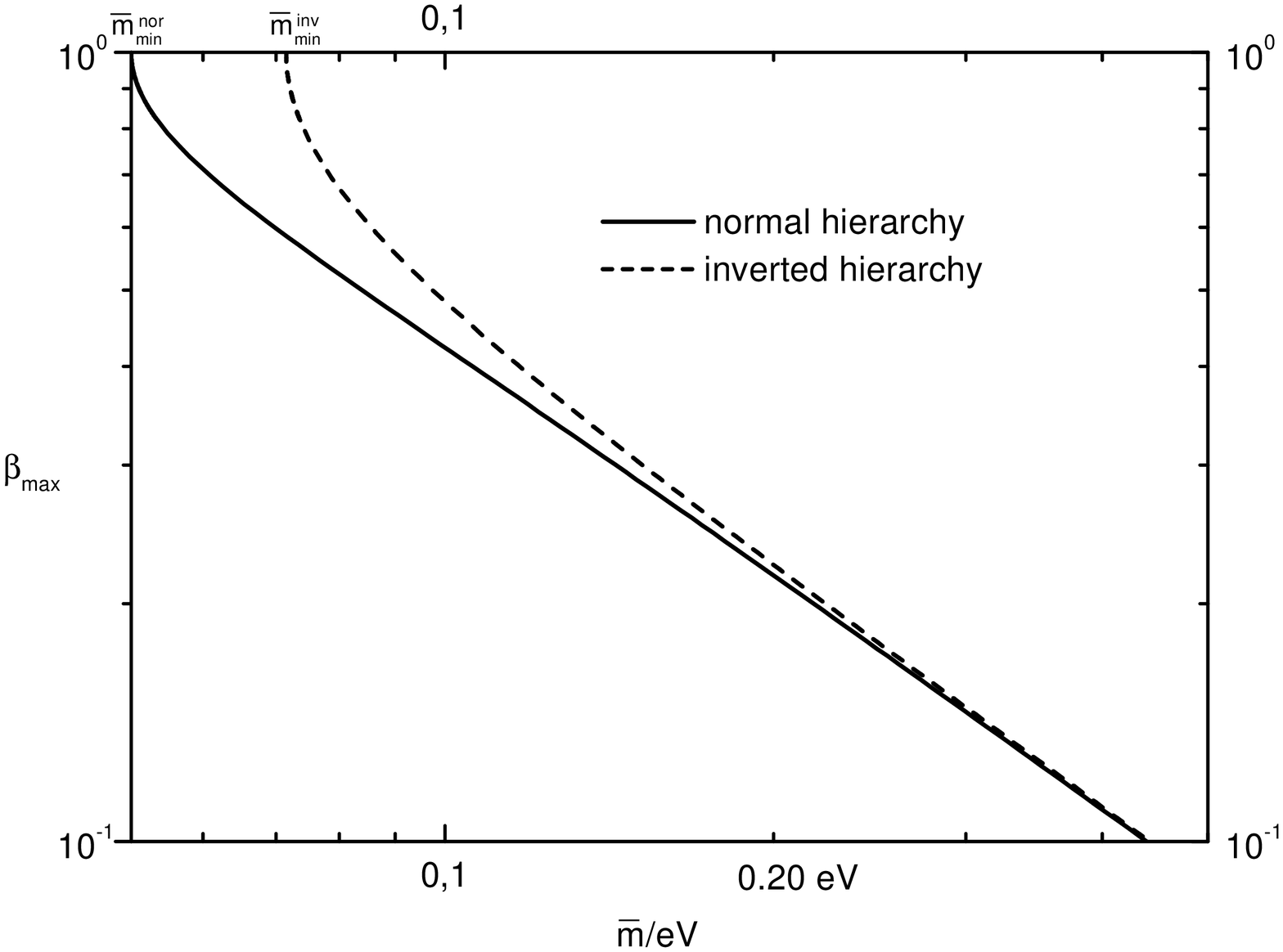,height=88mm,width=13cm}}
\caption{{\small The $C\!P$ asymmetry global suppression factor $\beta_{\rm max}$ 
for normal and inverted hierarchy.}}
\end{figure}
This implies that for increasing $\mb$ there is an overall suppression of the 
maximal baryon asymmetry in the whole $(\mt,M_1)$-plane \cite{di02}. In particular 
the lower limit on $M_1$ becomes more stringent.  

The factor $f(\mt,\mb) = 1$, for any value of $\mt$, if $\mb=\mb_{\rm min}$ 
($m_1=0$). In the case $\mb>\mb_{\rm min}$ ($m_1>0$) it vanishes for $\mt = m_1$ 
and grows monotonically to $1$ with increasing $\mt$ (cf.~Fig.~6). 
\begin{figure}
\centerline{\psfig{file=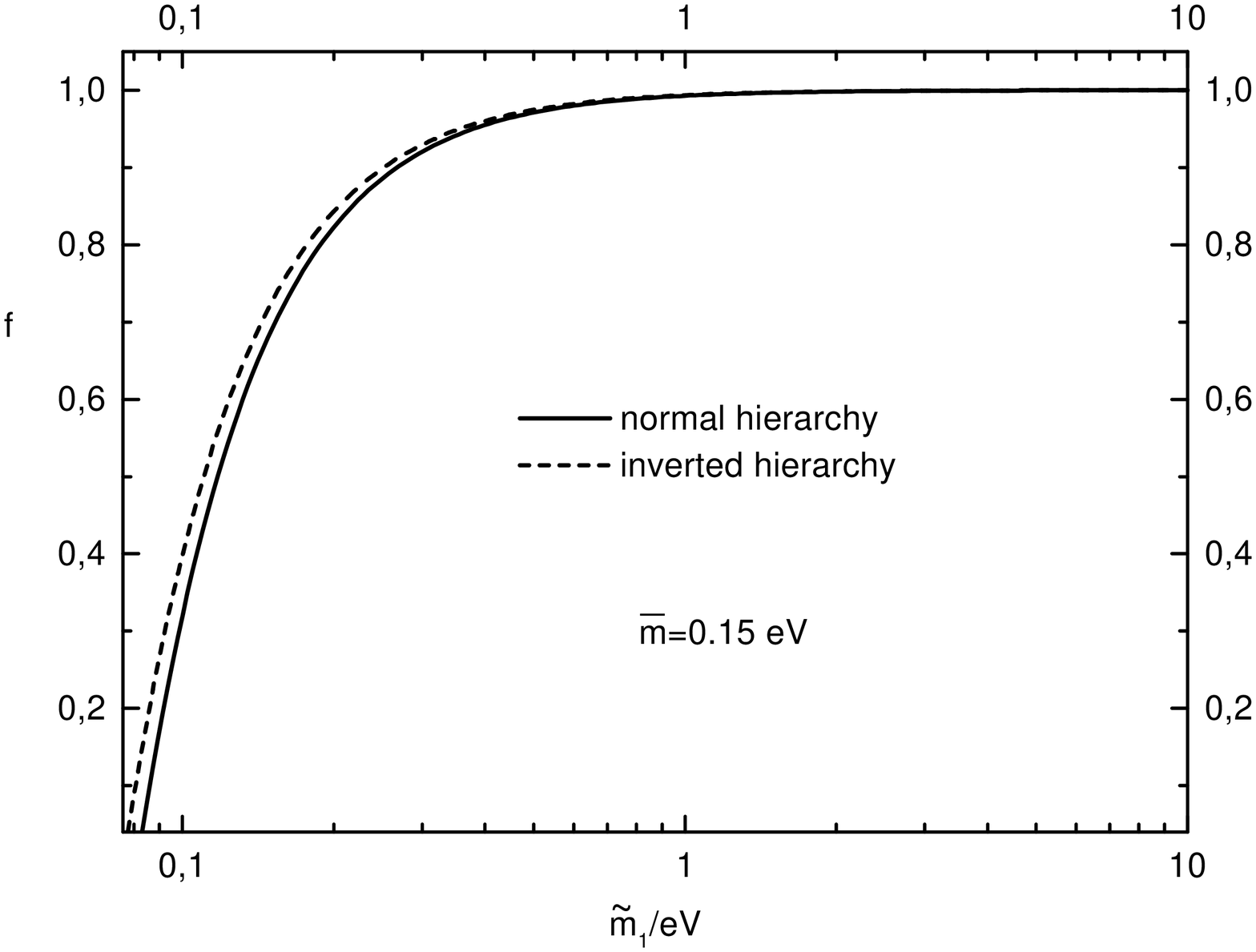,height=88mm,width=13cm}}
\caption{{\small The function $f(\widetilde{m}_1,\mb=0.15\,{\rm eV})$ for
normal (inverted) hierarchy. It is defined for 
$\mt\geq m_1\simeq 0.08\,(0.07)\,{\rm eV}$}.}
\end{figure}
Thus for $m_1 > 0$ the function $f$ gives a further suppression of the $C\!P$ 
asymmetry, in addition to the one from $\b_{\rm max} < 1$. This suppression 
is strong for $\mt \gtrsim m_1$ and disappears for $\mt \gg m_1$. 
Hence the decrease of the maximal $C\!P$ asymmetry for $\mb>\mb_{\rm min}$
shrinks the allowed region most at small $\mt \gtrsim m_1$ and at small $M_1$. 
Note that the difference between the allowed regions for normal and inverted
hierarchy is accounted for by the different values of $\b$ for a given value
of $\mb$. In the case of inverted hierarchy $\b$ is larger for any value of $\mt$ 
and  $\mb \geq \mb_{\rm min}^{\rm inv}$ (cf.~Figs.~5,6). The effect is maximal
for $\mb=\mb_{\rm min}^{\rm inv}$ where $\b^{\rm inv}=1$ while 
$\b^{\rm nor}\simeq 0.6$. For larger values of $\mb \gg \mb_{\rm min}^{\rm inv}$, 
and also $\mt \gg m_1^{\rm nor}$, the ratio $\b^{\rm inv}/\b^{\rm nor}$ becomes 
very close to $1$. This situation is realized when $\mb$ approaches its
upper bound. This explains why the upper bound on $\mb$ is
only slightly relaxed in the case of inverted hierarchy.
 
The second effect, which shrinks the allowed region when $\mb$ increases, is the
enhancement of washout processes. In \cite{bdp021} we showed how the total washout 
rate can be written as the sum of two terms, $(W - \D W) \propto \mt$ and  
$\D W \propto M_1\,\mb^2$. The first term is responsible for the reduction of 
the allowed region at large $\mt$. The second term leads to the boundary at 
large $M_1$. The combined effect shrinks the allowed region with increasing $\mb$ 
at large $M_1$ and at large $\mt$.
 
One can see how this second effect reduces the allowed region, independent of 
the maximal $C\!P$ asymmetry decrease, by comparing the two largest allowed regions
for normal and inverted hierarchy; they correspond to the two different values of 
$\mb_{\rm min}$ (cf.~Fig.~2 and Fig.~4). Since $\b=1$ in both cases, the entire
difference is due to the different washout effects. They are larger in the case of 
inverted hierarchy because $\mb_{\rm min}$ is about $\sim \sqrt{2}$ 
higher than in the normal hierarchy case. One can see how, for a fixed value of $\mt$,
the maximal value of $M_1$ is approximately halved in the inverted hierarchy case. 
Correspondingly, the maximal allowed value of $\mt$ is lower for inverted
hierarchy than for normal hierarchy.

In summary, within the theoretical uncertainties, leptogenesis cannot distinguish
between normal and inverted hierarchical neutrino mass patterns. However, our
new analysis confirms and strengthens the results of \cite{bdp021,bdp022} that
quasi-degenerate neutrino masses are strongly disfavored by leptogenesis, by 
putting the stringent upper bound of 0.12~eV on all neutrino masses.

\subsection{Stability of the bound}

The numerical results can be very well reproduced analytically \cite{prep}. 
This procedure is not only able to yield the correct value of $\mb_{\rm max}$ but also 
reveals some general features which in the numerical analysis may appear accidental.

For $\mb=\mb_{\rm max}$, at the peak value of maximal asymmetry, such that 
$\eta_B^{\rm max}=\eta_{B}^{CMB}$, one has
\begin{eqnarray} 
\mt|_{\rm max}&=&\mb_{\rm max}+{\cal O}
\left({m^2_{\rm atm}\over\mb^2_{\rm max}}\right)\;,\\ \label{M1max}
M_1|_{\rm max} & \simeq & 
1.6\times 10^{13}\,{\rm GeV}\,\left({0.2\,{\rm eV}\over \mb_{\rm max}}\right)^2\; .
\end{eqnarray}
The value of $\mb_{\rm max}$ is slightly different for normal and inverted hierarchy,
respectively,
\begin{eqnarray}
(\mb_{\rm max}^{\rm nor})^2 & = & (\mb_{\rm max}^0)^2
-{1\over 8}\,m^2_{\rm atm}+{\cal O}(m^4_{\rm atm}/\mb^4_{\rm max})\;, \\
(\mb_{\rm max}^{\rm inv})^2 & = & (\mb^0_{\rm max})^2
+{7\over 8}\,m^2_{\rm atm}+{\cal O}(m^4_{\rm atm}/\mb^4_{\rm max}) \;,
\end{eqnarray}
where $\mb_{\rm max}^0$ is the zero-th order approximation. This implies
\begin{equation}
(\mb_{\rm max}^{\rm inv})^2-(\mb_{\rm max}^{\rm nor})^2= 
m_{\rm atm}^2\,+{\cal O}(m^4_{\rm atm}/\mb^4_{\rm max})\; .
\end{equation}

Besides gaining more insight into the numerical results, the analytic procedure also 
allows to find the dependence of the bound on the involved physical parameters and 
to study in this way its stability.

Consider first the dependence on the experimental quantities $\eta_{B}^{CMB}$, 
$\D m^2_{\rm atm}$ and $\D m^2_{\rm sol}$. Since 
$\D m^2_{\rm sol}\ll \D m^2_{\rm atm}$, the dependence on $\D m^2_{\rm sol}$ is so 
small that it can be neglected, yielding $m_{\rm atm}\simeq\sqrt{\D m^2_{\rm atm}}$. 
The analytic procedure shows that 
$\mb_{\rm max}\propto (m_{\rm atm}^2/\eta_{B}^{CMB})^{1/4}$. 
From the numerical result, found for $\eta_{B}^{CMB}=3.6\times 10^{-10}$ and 
$m_{\rm atm}=m_0\simeq 0.051\,{\rm eV}$, and one then obtains in general
\begin{equation}\label{gen}
\mb_{\rm max}^0\simeq 
0.175\ {\rm eV}\ \left(6\times 10^{-10}\over \eta_B^{CMB}\right)^{1\over 4}
\ \left(m_{\rm atm}\over m_0 \right)^{1\over 2}\; .
\end{equation}

Using Eq.~(\ref{gen}) one immediately gets the central value of $\mb_{\rm max}$. 
Note also that for $\eta_{B}^{CMB}=10^{-10}$ and $m_{\rm atm}=m_0$, one has  
$\mb_{\rm max}^0\simeq 0.275\,\,{\rm eV}$. This is confirmed by the numerical 
results. We still find iso-lines $\eta_B^{CMB}=10^{-10}$ for $\mb=0.27\,{\rm eV}$, 
whereas  
this is not the case anymore for $\mb=0.28\,{\rm eV}$. From Eq.~(\ref{gen}) one
obtains as estimate for the relative error,
\begin{equation}
\d\mb_{\rm max}={1\over 4}\left(\d\eta_B^{CMB}+\d m^2_{\rm atm}\right)\;.
\end{equation}
According to Eq.~(\ref{CMB}) the $1\s$ standard error on $\eta_B^{CMB}$ is
about $15\%$ while $\delta m^2_{\rm atm}\simeq 25\%$ \cite{sk02}. We thus
obtain $\d \mb_{\rm max}\simeq 10\%$, which corresponds to the absolute error 
$\D \mb_{\rm max}\simeq 0.02\,{\rm eV}$. In the coming years the errors on 
$\eta_B^{CMB}$ and $m^2_{\rm atm}$ will be greatly reduced by the satellite 
experiments MAP \cite{map02} and Planck \cite{pla02}, and by the long baseline 
experiments Minos \cite{Minos} and CNGS \cite{CNGS}, respectively, and 
consequently the error on $\mb_{\rm max}$ will be considerably reduced. 

Another important question concerns the enhancement of the maximal $C\!P$ asymmetry 
when $\D M_{21} = M_2 - M_1$, where $M_1$ and $M_2$ are the masses of the heavy
neutrinos $N_1$ and $N_2$, becomes comparable to or smaller than $M_1$ itself. As
long as the mass splitting is larger than the decay widths, the enhancement is given 
by \cite{crv96,bp98},
\begin{equation}
\xi(x)={2\over 3}\,x\,\left[(1+x)\,\ln\left({1+x\over x}\right) 
-{2-x\over 1-x}\right]\; ,
\end{equation}
where $x=(M_2/M_1)^2$. Note, that $\xi$ approaches 1 for $x\gg 1$. The value of 
$\mb_{\rm max}$ increases like $\xi^{1/4}$ \cite{prep}, and it is therefore easy
to see how the bound on $\mb$ gets relaxed for small values of the mass difference
$\D M_{21}$. 

In Fig.~7 we have plotted the enhancement $\xi-1$ and the central value of 
$\mb_{\rm max}$, together with its $1\sigma$ limits, as function of 
$\Delta M_{21}/M_1$. 
\begin{figure}[t]
\centerline{\psfig{file=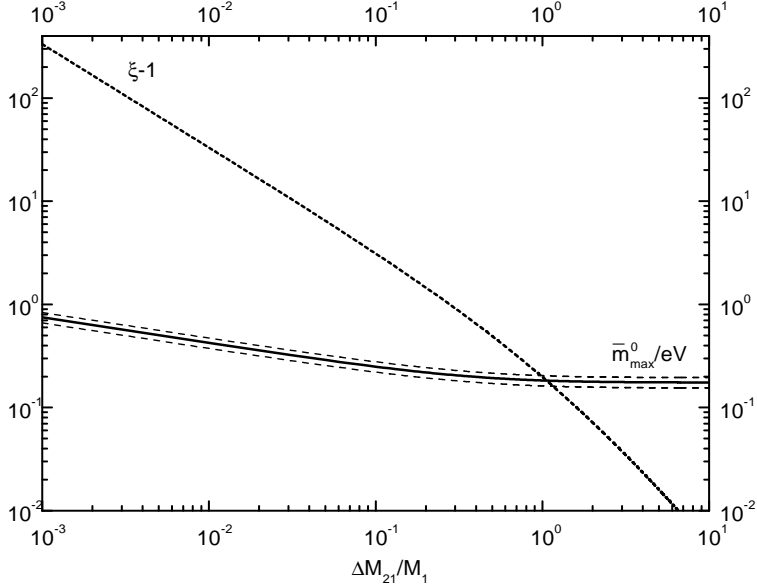,height=88mm,width=13cm}}
\caption{{\small The $C\!P$ asymmetry enhancement $\xi-1$ (short dashed line) and 
$\langle\mb_{\rm max}^0\rangle\pm \Delta\mb_{\rm max}^0$ for normal hierarchy 
(solid and dashed lines) as functions $\D M_{21}/M_1$.}}
\end{figure}
For $\D M_{21}/M_1\gtrsim 1$ the bounds (\ref{boundn}),(\ref{boundi}) are recovered. 
Only for values $\D M_{21}/M_1\lesssim 0.1$ the bound gets relaxed in an appreciable 
way. An increase of $\mb_{\rm max}$ by a factor $\sim 3$, allowing quasi-degenerate
neutrino masses of 0.4~eV, which could be detected with the KATRIN experiment 
\cite{kat01}, requires degeneracies $\D M_{21}/M_1,\D M_{31}/M_1 \lesssim 10^{-3}$.

In the regime $\D M_{21}\lesssim M_1$ also decays of $N_2$ have to be taken into
account. As we shall see in the next section, for larger mass splittings an asymmetry
generated in $N_2$ decays would be washed out before $T\sim M_1$, and it is then
sufficient to consider only $N_1$ decays. However, even for $\D M_{21}/M_1\lesssim 0.1$, 
it is easy to see that the effect of such an additional asymmetry on the bound is small 
compared to the effect of the $C\!P$ asymmetry enhancement described above.  The largest 
effect would be obtained for $\ve_2^{\rm max}=\ve_1^{\rm max}$ and 
$\widetilde{m}_2\ll\mt$, corresponding to a doubled heavy neutrino abundance without 
any washout enhancement. In this extreme case the bound is relaxed at most by a
factor $2^{1/4}\simeq 1.2$. Even for three degenerate neutrinos, with both 
$\D M_{21}/M_1 \ll 1$ and $\D M_{31}/M_1\ll 1$, the effect could relax the bound not 
more than by a factor $3^{1/4}\simeq 1.3$. Hence, the $C\!P$ enhancement represents the 
dominant effect and we can conclude that the bound on $\mb$ can only be evaded in case
of an extreme degeneracy among the heavy Majorana neutrinos.
 
A further important issue is the effect of supersymmetry on the bound. In this case
the maximal $C\!P$ asymmetry is about twice as large which could relax the bound by 
a factor $2^{1/4}\sim 1.2$. However, washout processes are also considerably  
enhanced \cite{plu98}. This effect goes into the opposite direction and is actually 
stronger, so that one can expect a slightly more stringent bound on $\mb$. A detailed 
calculation will be presented in \cite{prep}.

We conclude that the leptogenesis upper bound on neutrino masses is very stable.
The essential reason is that, at $\mb=\mb_{\rm max}$, the peak value 
$\eta_B^{\rm max} \propto 1/\mb_{\rm max}^4$. Hence, any variation of the final asymmetry
results into change of $\mb_{\rm max}$ which is almost one order of magnitude smaller.
The same argument applies also to the theoretical uncertainties. Although the various
corrections to the Boltzmann equations still remain to be calculated, we do not expect
a relaxation of $\mb_{\rm max}$ by more than $20\%$. In fact, we expect that the
corrections will go in the direction of lowering the prediction on the final asymmetry, 
which will make the bound on $\mb$ more stringent.

For particular patterns of neutrino mass matrices stronger bounds on the light neutrino 
masses can be obtained. For instance, one can study how the upper bound changes if $M_1$ 
is required to be smaller than some cut-off value $M_1^{\star}$. For 
$M_1^{\star}>M_1|_{\rm max}\simeq 10^{13}\,{\rm GeV}$ (cf.~(\ref{M1max})) the bound 
does not change. For smaller values of $M_1$ the bound becomes more stringent. For 
example, from Figs.~2 and 4 one can see that the cut-off
$M_1 < 5\times 10^{12}\,{\rm GeV}$ leads to the bound $\mb<0.15\,{\rm eV}$, which 
corresponds to $m_1 < 0.08\,{\rm eV}$. For a restricted mass pattern, and neglecting
$\D W\propto M_1\,\mb^2$ washout terms, less stringent bounds 
have been found in \cite{er02} for the same cut-off value of $M_1$.

\section{Dependence on initial conditions}

A very important question for leptogenesis, and baryogenesis in general, is
the dependence on initial conditions. This includes the dependence on the
initial abundance of heavy Majorana neutrinos, which has been studied in 
detail in \cite{bdp021}, and also the effect of an initial asymmetry which
may have been generated by some other mechanism. In the following we shall
study the efficiency of the washout of a large initial asymmetry by heavy
Majorana neutrinos. 

For simplicity, we neglect the small asymmetry generated through the
$C\!P$ violating interactions of the heavy neutrinos, i.e. we set $\ve_1=0$.
The kinetic equation (\ref{lg2}) for the asymmetry then becomes
\begin{equation}\label{kew}
{dN_{B-L}\over dz} = -W\,N_{B-L} \;,
\end{equation}
where $-N_{B-L}$ is the number of lepton doublets per comoving volume.
The final $B-L$ asymmetry is then given by
\begin{equation}\label{final}
N_{B-L}^{\rm f} = \o(z_{\ri}) N_{B-L}^{\rm i}\;,
\end{equation}
with the washout factor
\begin{equation}
\o(z_{\ri}) = e^{-\int_{z_{\rm i}}^{\infty}\,dz\,W(z)}\;.
\end{equation} 

In Eq.~(\ref{kew}) $W(z)=\G_W(z)/H(z)z$ is the rescaled washout rate, where
$H(z)$ is the temperature-dependent Hubble parameter. $\G_W$ receives contributions
from inverse decays ($\G_{ID}$), $\D L=1$ processes ($\G_{\phi,t}$, $\G_{\phi,s}$)
and $\D L=2$ processes ($\G_N$, $\G_{N,t}$) (cf.~\cite{bdp021}),
\begin{equation}
\G_W={1\over 2}\G_{ID}+2\left(\G_N^{(l)}+ \G_{N,t}^{(l)}\right) +
2\G_{\phi,t}^{(l)}+ {n_{N_1}\over n_{N_1}^{\rm eq}}\G_{\phi,s}^{(l)}\;.
\end{equation}
The inverse decay rate is given by
\begin{equation}
\G_{ID}={n_{N_1}^{\rm eq}\over n_{l}^{\rm eq}}\,\G_D\;, \quad
\G_D= {1\over 8\p} \left(h^\dg h\right)_{11} M_1 {K_1(z)\over K_2(z)}\;,
\end{equation}
where $n_{N_1}^{\rm eq}$ and $n_{l}^{\rm eq}$ are the equilibrium number densities
of heavy neutrinos and lepton doublets, respectively, and $K_{1,2}(z)$ are 
modified Bessel functions of the third kind. 
The quantities $\G_{i}^{(X)}$ are thermally averaged reaction rates per particle $X$.
They are related by $\G_i^{(X)}=\g_i/n_{X}^{\rm eq}$ to the reaction densities 
$\g_i$ \cite{lut92}.
Our calculations are based on the reduced cross sections given in ref.~\cite{plu98}.

It is very instructive to consider analytical approximations to the various washout
contributions. Both, the inverse decay rate and the resonance part of $\G_N^{(l)}$
(cf.~\cite{bdp021}) are proportional to $K_1(z)$,
\begin{equation}
\G_W^{(1)} = {1\over 2}\G_{ID} + 2\G_{N,res}^{(l)}
= {1\over 16\p\z(3)}\left(h^\dg h\right)_{11} M_1 z^2 K_1(z)\;.
\end{equation}
The integral in Eq.~(\ref{kew}) can be analytically performed. The corresponding 
washout factor can be written in the form
\begin{eqnarray}
\o^{(1)}(z_{\ri}) = \exp{\left\{-{1\over 2\z(3)}{\mt \over m_*}\left({3\p\over 2} 
+ z_i^3K_2(z_i) 
-{3\p\over 2} z_i\left(K_2(z_i)L_1(z_i)+L_2(z_i)K_1(z_i)\right)\right)\right\}} ,
\end{eqnarray}
where $m_*$ is the equilibrium mass (\ref{mequ}), and $L_{1,2}(z)$ are modified
Struve functions \cite{bateman}. Rather accurate approximations are, for small and 
large values of $z_{\ri}$ respectively,
\begin{equation}\label{wash1}
\o^{(1)}(z_{\ri}) =
\left\{ \begin{array}{ll}  
\exp{\left\{-{1\over 2\z(3)}{\mt \over m_*}
\left({3\p\over 2} - {1\over 3}z_i^3 + {\cal O}(z_i^5)\right)\right\}}\;, 
\; z_{\ri} < 1 & \\ [2ex]
\exp{\left\{-{1\over 2\z(3)}{\mt \over m_*}\sqrt{\p\over 2z_i}e^{-z_i}
\left(z_i^3 + {23\over 8}z_i^2 + {537\over 128}z_i + {2253\over 1024} 
+ {\cal O}({1\over z_i})\right)\right\}}\; , \; z_i > 1 \; . 
\end{array}\right.
\end{equation} 

The non-resonant contribution of $N_1$ exchange to the washout is proportional to
$\mb^2$,
\begin{eqnarray}
\G_W^{(2)} &=& 2 \left(\G^{(l)}_{N,nonres} + \G^{(l)}_{N,t}\right) \NO\\
&=& {1\over \p^3\z(3)}{M_1^3 \mb^2\over v^4}{1\over z^3}\;,
\end{eqnarray}
which yields the washout factor
\begin{equation}\label{wash2}
\o^{(2)}(z_{\ri}) = \exp{\left\{- {8 \over \p^2\z(3)} {M_1 \mb^2\over m_*v^2}
{1\over z_i}\right\}}\;.
\end{equation} 

Finally, we have to consider $N_1$-top scatterings. The rate is dominated by 
t-channel Higgs exchange if the infrared divergence is cut off by a Higgs mass
$m_{\phi} \sim 1$~TeV~$\ll T$. In terms of the reduced cross section one has  
(cf.~\cite{plu98}),
\begin{equation}
\G^{(l)}_{\phi,t} = {M_1 z^2\over 96\p^2\z(3)}\int_1^\infty dx \sqrt{x}
\hat{\s}_{\phi,t}(x) K_1(z\sqrt{x})\;.
\end{equation}
For small and large values of $z_{\ri}$, respectively, analytic expressions are
given by
\begin{equation}\label{wash3}
\o^{(3)}(z_{\ri}) =  
\left\{\begin{array}{ll}
\exp{\left\{-{\a_u\over 2\z(3)}{\mt \over m_*}
\left(\ln{(4 a_{\phi})} - 1\right)\right\}}\;, z_{\ri} < 1 & \\ [2ex]
\exp{\left\{-{\a_u\over 2\z(3)}{\mt \over m_*}\sqrt{2 z_i\over \p }e^{-z_i}
\left(\ln{\left({a_{\phi}\over z_i^2}\right)}\left({11\over 8}+z_i\right)
+ {5\over 8} - z_i\right)\right\}} \;, z_i > 1 \; , \end{array} \right.
\end{equation} 
where $\a_u = g_t^2/(4\p)$ and $a_{\phi}=M_1^2/m_{\phi}^2$.

\begin{figure}
\centerline{\psfig{figure=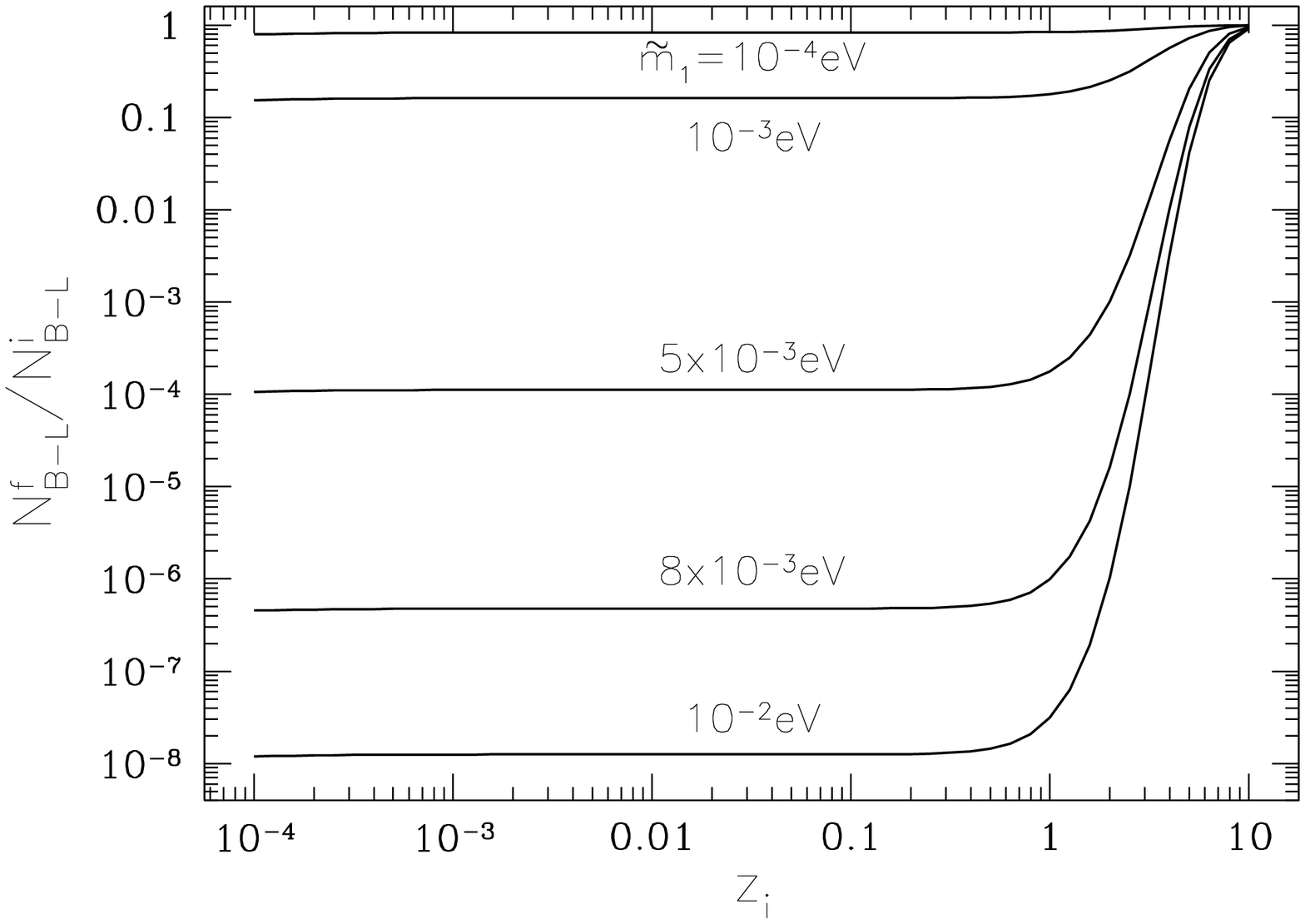,width=13cm}}
\caption{Washout factor as function of the initial temperature $z_i=M_1/T_{\ri}$
for different values of $\mt$ and $M_1=10^8$~GeV; $N_1$-top scatterings
are neglected.}
\label{fig:wash1}
\end{figure}
\begin{figure}
\centerline{\psfig{figure=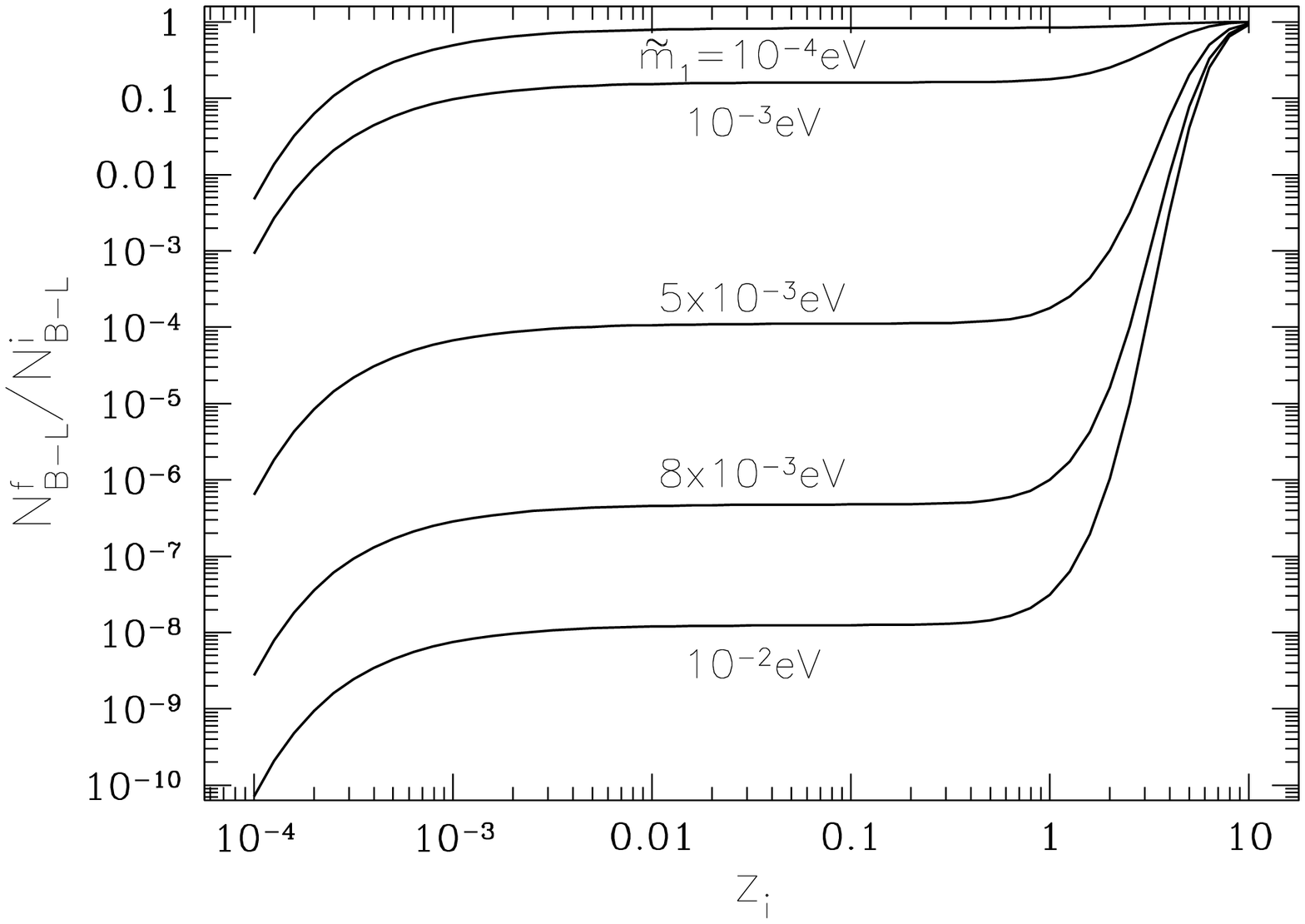,width=13cm}}
\caption{Washout factor as function of the initial temperature $z_i=M_1/T_{\ri}$
for different values of $\mt$ and $M_1=10^{10}$~GeV; $N_1$-top scatterings
are neglected.}
\label{fig:wash2}
\end{figure}
\begin{figure}
\centerline{\psfig{figure=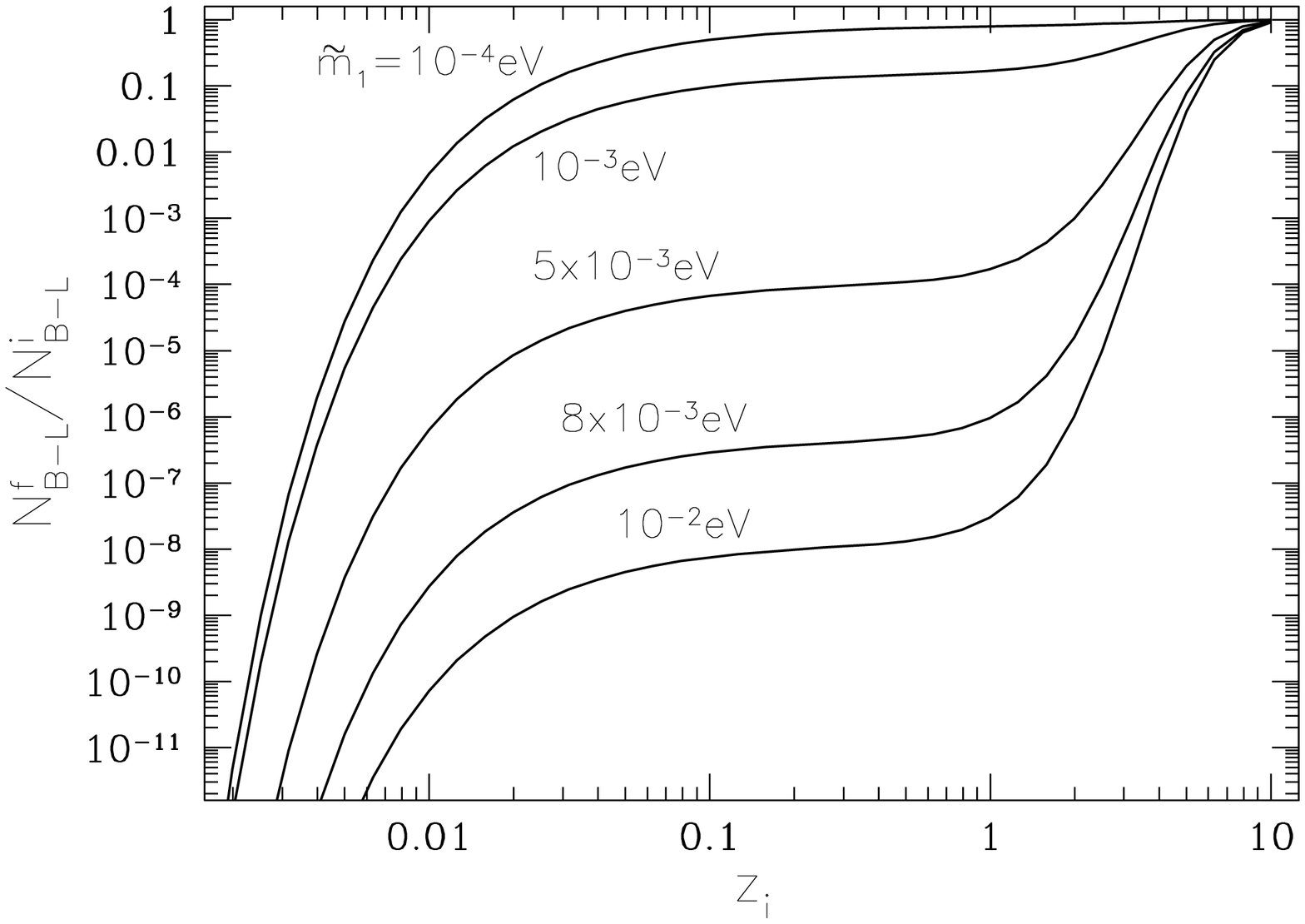,width=13cm}}
\caption{Washout factor as function of the initial temperature $z_i=M_1/T_{\ri}$
for different values of $\mt$ and $M_1=10^{12}$~GeV; $N_1$-top scatterings
are neglected.}
\label{fig:wash3}
\end{figure}
\begin{figure}
\centerline{\psfig{figure=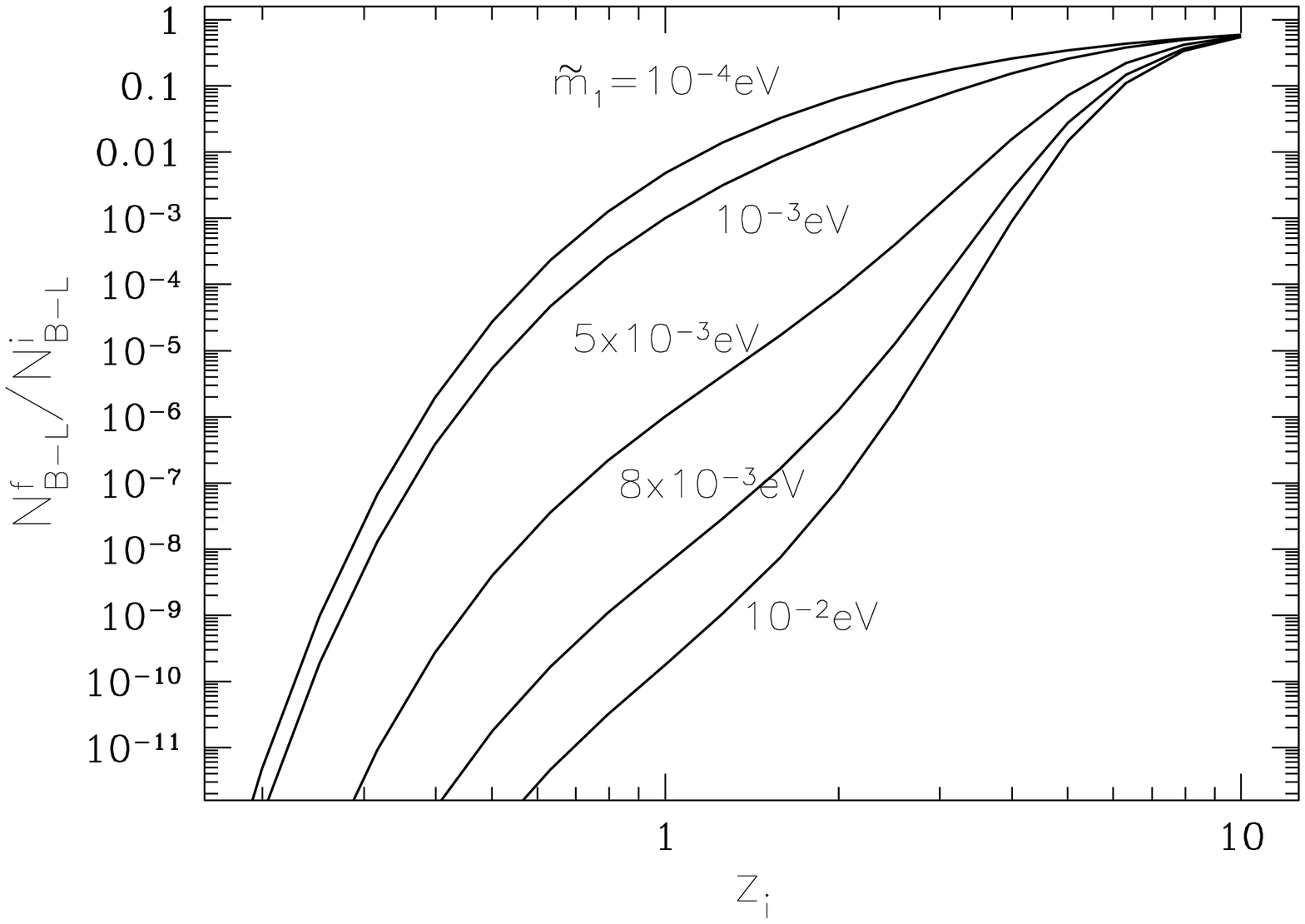,width=13cm}}
\caption{Washout factor as function of the initial temperature $z_i=M_1/T_{\ri}$
for different values of $\mt$ and $M_1=10^{14}$~GeV; $N_1$-top scatterings
are neglected.}
\label{fig:wash4}
\end{figure}

The total washout factor
\begin{equation}
\o(z_{\ri}) = {N_{B-L}^{\rf} \over N_{B-L}^{\ri}} 
= \o^{(1)}(z_{\ri}) \o^{(2)}(z_{\ri}) \o^{(3)}(z_{\ri})
\end{equation}
depends exponentially on the parameters $\mt$ ($\o^{(1)}$,$\o^{(3)}$) and 
$M_1\mb^2$ ($\o^{(2)}$). For not too large $M_1$ and not too small $z_{\ri}$ 
(cf.~Figs.~(\ref{fig:wash1})-(\ref{fig:wash3})), $\o^{(2)} \simeq 1$ whereas
$\o^{(1)}$ reaches a plateau for $z_{\ri} \leq 1$ at
\begin{equation}
\o^{(1)}(z_{\ri}) \simeq 
\exp{\left(-{3\p\over 4\z(3)}{\mt \over m_*}\right)}\;.
\end{equation}
At smaller values of $z_{\ri}$, and correspondingly higher temperatures $T_{\ri}$,
eventually $\o^{(2)}$ decreases rapidly. When $T_{\ri}$ reaches 
$M_2$, the mass of $N_2$, a new plateau will be reached. The larger $M_1$, the 
larger the value
of $z_{\ri}$ where the decrease of $\o^{(2)}$ sets in. This behaviour is clearly
visible in Figs.~(\ref{fig:wash1})-(\ref{fig:wash3}). At very large $M_1$,
the decrease of $\o^{(2)}$ is effective already at large values of $z_{\ri}$
(cf.~Fig.~(\ref{fig:wash4})).

\begin{figure}
\centerline{\psfig{figure=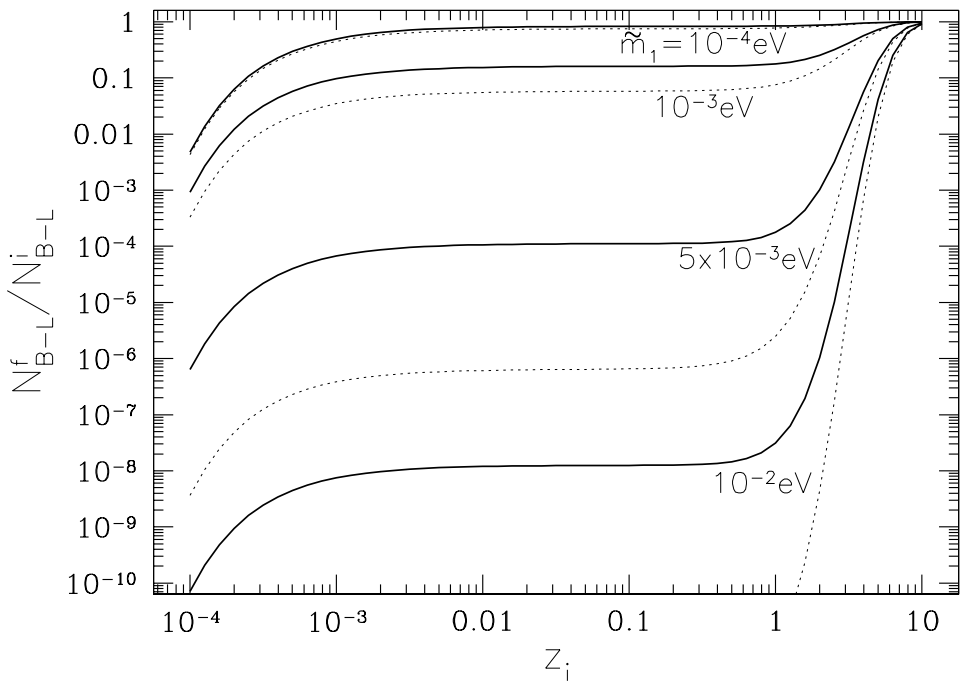,width=13cm}}
\caption{Comparison of the washout factors as function of $z_i=M_1/T_{\ri}$ without 
(full line) and with (dashed line) $N_1$-top scatterings; $M_1=10^{10}$~GeV.}
\label{fig:comp}
\end{figure}

The factor $\o^{(3)}$ is very sensitive to the value of $a_{\phi}$, i.e. the choice
of the infrared cutoff $m_{\phi}$. For $m_{\phi} = 1$~TeV, $\o^{(3)}$ significantly
improves the washout of $\o^{(1)}\o^{(2)}$, but it does not change the qualitative
picture. This is illustrated by Fig.~(\ref{fig:comp}) where the cases with and
without $N_1$-top scatterings are compared. On the other hand, for
$m_{\phi} \sim M_1$, $\o^{(3)}$ is always negligible compared to $\o^{(1)}$.
The issue of the correct choice of the infrared cutoff is theoretically not yet 
settled. There is a corresponding, though less important uncertainty in the 
generation of the baryon asymmetry for small values of $\mt$ \cite{bcx00}.
The washout factors $\o^{(1)}\o^{(2)}$ shown in 
Figs.~(\ref{fig:wash1})-(\ref{fig:wash4}) can be regarded as conservative upper
bounds on the full washout factors $\o=\o^{(1)}\o^{(2)}\o^{(3)}$.

It is remarkable that the washout of an initial asymmetry at $z_{\ri} \sim 1$, i.e. 
$T_{\ri} \sim M_1$, becomes very efficient for $\mt\geq m_*\simeq 10^{-3}$~eV.
Since the efficiency increases exponentially with increasing $\mt$, already at 
$\mt = 5\times10^{-3}$~eV one has $\o(z_{\ri}=1) < 10^{-4}$. Hence, for neutrino
masses of order or larger than $\sqrt{\D m^2_{\rm sol}}$, $\D L=1$ processes are very 
likely to erase any previously generated baryon asymmetry to a level below the 
asymmetry produced by leptogenesis. As shown in \cite{bdp021}, for these values of
$\mt$ the final asymmetry is also independent of the initial $N_1$ abundance. Hence,  
a complete independence of initial conditions is achieved.

\section{Summary}

We have extended our previous work on the minimal version of thermal leptogenesis
where interactions of $N_1$, the lightest of the heavy Majorana neutrinos, are the
dominant source of the baryon asymmetry. Based on the seesaw mechanism,
we have derived an improved upper bound on the $C\!P$ asymmetry $\ve_1$, which
depends on $M_1$, the mass of $N_1$, the light neutrino masses $m_1$ and $m_3$,
and the effective neutrino mass $\mt$. Given the two mass splittings $\D m^2_{\rm atm}$
and $\D m^2_{\rm sol}$, the neutrino masses $m_1$ and $m_3$ can depend on the absolute 
neutrino mass scale $\mb$ in two ways, corresponding to normal and inverted
mass hierarchy, respectively.

From the numerical solution of the Boltzmann equations we have obtained an upper
bound on all light neutrino masses of 0.12~eV, which holds for normal as well as
inverted neutrino mass hierarchy. This is about a factor of two below the recent
upper bound of $0.23$~eV obtained by MAP \cite{map}. 
The leptogenesis bound is remarkably stable with respect to 
changes of $\eta_B^{CMB}$, $\D m^2_{\rm atm}$, the effect of supersymmetry, and 
theoretical uncertainties of $\eta_B^{\rm max}$. Quasi-degenerate neutrinos are only
allowed if the $C\!P$ asymmetry is strongly enhanced by a degeneracy of the heavy
Majorana neutrinos. For instance, in order to relax the upper bound to 0.4~eV,
degeneracies $\D M_{21}/M_1,\D M_{31}/M_1 \lesssim 10^{-3}$ are required.  

We have also studied the washout of a large, pre-existing $B-L$ asymmetry. It is 
very interesting that a washout by several orders of magnitude takes place at
temperatures $T$ close to $M_1$, if the effective neutrino mass $\mt$ is larger
than the equilibrium mass $m_* \simeq 10^{-3}$~eV. All memory 
of the initial conditions is then erased.

We conclude that for neutrino masses in the range from $10^{-3}$~eV to 0.1~eV 
leptogenesis naturally explains the observed baryon asymmetry, independent of possible 
other pre-existing asymmetries. It is very remarkable that the data on solar and 
atmospheric neutrinos indicate neutrino masses precisely in this range. \\

\noindent
{\bf Acknowledgments}\\
P.D.B. was supported by the EU Fifth Framework network ``Supersymmetry
and the Early Universe" (HPRN-CT-2000-00152).

\end{document}